\documentclass[11pt,letterpaper]{article}
\usepackage[margin=1in]{geometry}
\usepackage{amsmath,amssymb,amsthm,mathtools,booktabs}
\usepackage[colorlinks=true,linkcolor=blue,citecolor=blue,urlcolor=blue]{hyperref}

\usepackage{graphicx}
\usepackage{tikz}
\usetikzlibrary{arrows.meta,fit,backgrounds,calc}

\usepackage{algorithm}
\usepackage{algpseudocode}
\usepackage{appendix}
\usepackage{xcolor}
\usepackage{multirow}
\usepackage{bm}
\usepackage{etoolbox}

\definecolor{mydustyblue}{HTML}{104E8E}  

\hypersetup{
    linkcolor=mydustyblue,     
    citecolor=mydustyblue,   
    urlcolor=mydustyblue,     
    filecolor=mydustyblue     
}

\usepackage{authblk}

\newtheorem{theorem}{Theorem}[section]
\newtheorem{lemma}[theorem]{Lemma}
\newtheorem{proposition}[theorem]{Proposition}
\newtheorem{corollary}[theorem]{Corollary}

\newtheorem{claim}[theorem]{Claim}
\theoremstyle{definition}
\newtheorem{definition}[theorem]{Definition}
\theoremstyle{remark}
\newtheorem{remark}[theorem]{Remark}
\newtheorem{observation}[theorem]{Observation}

\newcommand{\vol}{\operatorname{vol}}

\newcommand{\RQ}{\mathcal{R}}
\newcommand{\eps}{\varepsilon}

\newcommand{\one}{\mathbf{1}}

\newcommand{\TildeO}{\widetilde O}

\title{Faster Cluster Decomposition with Stronger Bi-Criteria Guarantee}
\author[1,2]{He Sun\thanks{Email: hsun@siat.ac.cn}}
\author[1]{Yong Zhang\thanks{Email: zhangyong@siat.ac.cn}}
\author[1]{Yacong Zhou\thanks{Email: yacong.zhou96@gmail.com}}

\affil[1]{Shenzhen Institutes of Advanced Technology, Chinese Academy of Sciences}

\affil[2]{School of Informatics, University of Edinburgh}

\begin{document}

\pagenumbering{gobble}

\maketitle

\begin{abstract}
Partitioning the vertex set of a graph  $G=(V,E)$
into clusters is a classical problem in computer science. As vertices within a cluster should be well connected while cuts between different clusters should be sparse, Oveis Gharan and Trevisan~(SODA 2014, \cite{GT14}) proposed to measure the partition quality using both the \emph{inner} conductance and the \emph{outer} conductance, i.e., the \emph{bi-criteria guarantee}. Specifically, when partitioning $V$ into vertex-disjoint subsets $P_1,\ldots, P_k$, every subgraph induced by  $P_i$ should have high conductance, while the conductance of  $P_i$  in $G$ should be low. They studied the existence of such a partition, and designed a polynomial-time algorithm that finds a partition with  the  bi-criteria guarantee.
However, for more than a decade, their algorithm remained the only one that 
achieves this bi-criteria guarantee on the output clusters.

In this paper, we study faster  cluster decomposition algorithms under the same setup as  \cite{GT14}, and prove the following results: 
\begin{enumerate}
    \item We make their existential result algorithmic: Our algorithms run in \emph{nearly-linear time}, and achieve the same bi-criteria guarantee  as the one in \cite{GT14}  up to constant factors, including the asymptotically tight dependence on $k$.
    \item We prove that their existential result holds \emph{algorithmically} under a much weaker condition  on the initial disjoint subsets.  To achieve this, we develop a general algorithm framework that boosts the approximation guarantee of  a solution:  given an $\alpha$-approximate solution to the $k$-way conductance problem, if
$\rho_G(k+1)/\rho_G(k)=\Omega\left(\alpha/\varepsilon\right)$, then a $(1+\varepsilon)$-approximate solution can be obtained using only $O(k(1+\log\alpha))$ minimum $s$-$t$ cut computations.
    \item  We present a \emph{nearly-linear time} algorithm that finds clusters with the bi-criteria guarantee. Our algorithm has the same guarantee as the  one in \cite{GT14}, which, for any unweighted graph of $n$ vertices and $m$ edges, runs in $\TildeO(knm^2)$ time.   

\end{enumerate}

\end{abstract}

\newpage

\tableofcontents

 \newpage

\pagenumbering{arabic}

\section{Introduction}

 Partitioning a graph into disjoint sets of vertices under different constraints is a fundamental problem in computer science, and has been extensively studied under the  names of graph partitioning, clustering  and  community detection. While this problem naturally arises in different disciplines including theoretical computer science, machine learning, and social network analysis~\cite{ShiMalik00,FortunatoHric16}, identifying the right notion of \emph{clusters} is far from straightforward. 
 While    a good measure of clustering quality often involves optimizing several objective functions, 
 Kannan, Vempala and Vetta~\cite{KVV04} proved that  optimizing a single function can lead to an undesirable clustering. Based on this, they  propose a bi-criteria measure, under which they measure the quality of a $k$-clustering based on the conductance of the subgraphs induced by every cluster, and the edges between the clusters. Oveis Gharan and Trevisan~\cite{GT14} further proposed to use the \emph{inner} and \emph{outer} conductance of clusters to measure the quality of a clustering; i.e., a good cluster $S$ in $G=(V,E)$ should have the property that (i) the vertices inside $S$  are  well connected, and (ii) there is a sparse cut between $S$ and $V\setminus S$. This  work has inspired many algorithmic studies that find clusters based on their inner and outer conductance, e.g.~\cite{CPS15,DPRS19,Peng20,GKLMS21,
SP23,PSZ15,LMS23,HFP24,SZPH26}.


In this paper we study efficient algorithms that partition an input graph into clusters with respect to the inner and outer conductance of the clusters. For any weighted graph $G=(V,E,w)$ of $n$ vertices and set $S\subset V$, the \emph{outer conductance} of $S$ is defined by \[
\phi_G(S)
\triangleq
\frac{w(S,V\setminus S)}{\vol_G(S)},
\]
where  $w(S,V\setminus S)$ is the cut value between $S$ and $V\setminus S$, and $\vol_G(S)$
denotes the total degree of vertices
in $S$. The \emph{conductance} of $G$ is defined by 
\[
\phi(G)\triangleq
\min_{\substack{
S\subseteq V\\
0<\vol_G(S)\leq\vol_G(V)/2
}}
\phi_G(S).
\]
Generalizing the conductance     for multiple disjoint subsets, we define the \emph{$k$-way expansion} of $G$ by 
\[
\rho_G(k) \triangleq
\min_{\substack{
A_1,\ldots,A_k\subseteq V\\
A_i\cap A_j=\emptyset, i\neq j\\
\vol_G(A_i)>0 
}}
\max_{1\leq i\leq k}\phi_G(A_i).
\]
For any subset $S\subseteq V$ with a positive volume,
let $G\{S\}$ denote the
\emph{degree-preserving induced subgraph}
obtained from $G[S]$ (the subgraph induced by $S$) by adding a self-loop of weight $w(v,V\setminus S)$ at each
vertex $v\in S$, so that every vertex
maintains its original degree in $G$.
The \emph{inner conductance}
of $S$ is defined as $\phi(G\{S\})$. Hence, by definition $S$ is a cluster if $\phi(G\{S\})$ is large and $\phi_G(S)$ is small.

\subsection{Our Results}

 Throughout the paper, we assume that every vertex of $G$ has positive degree,  the weight of every edge of $G$ is an integer, and the ratio between the maximum weight of any edge and the minimum one is upper bounded by a polynomial in $n$. 
We first study the cluster decomposition problem for any graph $G$ that has a gap between $\rho_G(k+1)$ and $\rho_G(k)$. Oveis Gharan and Trevisan~\cite{GT14} showed that, if $\rho_G(k+1)>(1+\eps)\rho_G(k)$ and $A_1,\dots,A_k$ are $k$ disjoint subsets satisfying
$\max_{1\leq i\leq k}\phi_G(A_i)=\rho_G(k)$, then
one can construct, from $A_1,\dots,A_k$, $k$
pairwise disjoint subsets $B_1,\ldots,B_k$
satisfying
\[\phi_G(B_i)\leq\rho_G(k) \qquad\text{and}\qquad
\phi(G\{B_i\}\footnote{
Although Oveis Gharan and Trevisan~\cite{GT14}
stated their inner conductance guarantees
in terms of $\phi(G[S])$, their proofs   actually establish
the stronger bounds for $\phi(G\{S\})$.
Moreover, this distinction is essential
for their spectral analysis: the
induced-subgraph formulation of
\cite[Lemma~1.1]{GT14} is incorrect in general,
whereas the corresponding degree-preserving
statement holds.
See Section~\ref{apx:counterexamples}
for a counterexample and further discussion.
} )\geq
\frac{\eps\rho_G(k+1)}{7},\]
for every $i\in[k]$.
Moreover, these subsets can be extended
to a $k$-partition
$V=P_1\sqcup\cdots\sqcup P_k$
satisfying
\[\phi_G(P_i)\leq k\rho_G(k) \qquad\text{and}\qquad
\phi(G\{P_i\})\geq
\frac{\eps\rho_G(k+1)}{14k},\]
for every $i\in[k]$.
They also showed that some separation
between $\rho_G(k)$ and $\rho_G(k+1)$
is necessary in general, and that the
linear dependence on $k$ in the partitioning
guarantees is asymptotically tight.
However, their results only show the existence of the subsets.

We first prove that, under the same assumption of $\rho_G(k+1)>(1+\eps)\rho_G(k)$, both the $k$ disjoint subsets $B_1,\ldots, B_k$ and the $k$-way partition $V=P_1\sqcup\cdots\sqcup P_k$ can be constructed in nearly-linear time, and the constructed subsets have asymptotically  the same inner and outer conductance guarantee as theirs. 
Moreover, the initial subsets need not
be optimal: our algorithms work with any
$k$ disjoint subsets with conductance
 at most $p$, provided that
$\rho_G(k+1)\geq(1+\eps)p$. 

\begin{theorem} 
\label{thm:main}
Let $k\geq2$, $0<\eps\leq1$, and $p>0$.
Suppose that $A_1,\ldots,A_k$ are pairwise disjoint,
nonempty, positive-volume subsets of $V$ satisfying
\[
\phi_G(A_i)\leq p
\qquad\text{for every }i\in[k],
\]
and $\rho_G(k+1)\geq(1+\eps)p$. Then,
\begin{itemize}
    \item [\textup{(i)}] there is a randomized algorithm (Algorithm \ref{alg:fair-trimming}) that
    runs in $\TildeO(m/\eps)$ time\footnote{Throughout the paper the notation  $\TildeO(\cdot)$ hides a factor of $\log^cn$ for a positive constant $c$.} and outputs
    $k$ pairwise disjoint subsets $B_1,\dots,B_k$
    (with $B_i\subseteq A_i$ for all $i\in[k]$)
    such that it holds for every $i\in[k]$ that 
    \[
    \phi_G(B_i)\leq
    \left(1+\frac{\eps}{16}\right)p
    \qquad\text{and}\qquad
    \phi(G\{B_i\})\geq
    \frac{\eps\rho_G(k+1)}{9};
    \]
    \item [\textup{(ii)}] there is a randomized algorithm (formed by Algorithms \ref{alg:fair-trimming} and \ref{alg:fair-attachment})
    that runs in
    $\TildeO(m(\eps^{-1}+\log^2 k))$ time
    and outputs a partition
    $V=P_1\sqcup\cdots\sqcup P_k$
    such that it holds for every $i\in[k]$ that 
    \[
    \phi_G(P_i)\leq4kp
    \qquad\text{and}\qquad
    \phi(G\{P_i\})\geq
    \frac{\eps\rho_G(k+1)}{53k}.
    \]
\end{itemize}
\end{theorem}
Taking $p=\rho_G(k)>0$ and letting
$A_1,\ldots,A_k$ be an optimal  family of  $k$ pairwise disjoint subsets,
Theorem~\ref{thm:main} recovers   the two
existential guarantees of \cite[Theorem 1.6]{GT14}
up to constant factors.
In particular, our algorithms preserve
the asymptotically tight dependence on
$k$ while requiring only nearly-linear
additional time once the initial subsets
are given.

To achieve this result, we separate the objective   of finding
low-conductance subsets from that of
turning them into internally expanding
clusters. This distinction is important, because
finding an optimal low-conductance cut
is already \textsf{NP}-hard   for $k=2$ and computing an optimal  family of 
$k$ disjoint subsets cannot generally be
used as an efficient initialization step.
To overcome this, our algorithms start from
any approximate solution and perform
the expander construction as a separate procedure that runs in nearly-linear time. 
More concretely, suppose that an
$\alpha$-approximation algorithm for
$\rho_G(k)$ returns $k$ disjoint subsets
$A_1,\ldots,A_k$.
Setting
$p=\max_{i\in[k]}\phi_G(A_i)
\leq\alpha\rho_G(k)$,
Theorem~\ref{thm:main} applies whenever $\rho_G(k+1)\geq
(1+\eps)\alpha\rho_G(k)$.
It then produces a $k$-way partition
with outer conductance
$O(k\alpha\rho_G(k))$
and   inner conductance
$\Omega(\eps\rho_G(k+1)/k)$.
In other words,   our result reduces the problem of cluster decomposition   to the one
of finding $k$ disjoint low-conductance
subsets.

Notice that \cite[Theorem 1.6]{GT14} constructed the subsets with the  bi-criteria guarantee from  an \emph{optimal} family of $k$ pairwise disjoint  subsets $A_1,\ldots, A_k$ corresponding to $\rho_G(k)$, which is \textsf{NP}-hard to compute. This raises an important question: can we construct the subsets with the bi-criteria guarantee without knowing the optimal $A_1,\ldots, A_k$? To address this, we study the  trade-off  between the ``quality'' of the initial $k$ disjoint subsets and the value of $\rho_G(k+1)/\rho_G(k)$.  
In particular, we give a general
approximation-boosting procedure that
turns an arbitrary $\alpha$-approximate
solution into a $(1+\eps)$-approximate
solution for $\rho_G(k)$, provided that
the gap is sufficiently large.

\begin{theorem} 
\label{thm:epsapprox}
Let $G=(V,E,w)$ be a weighted graph, $\eps>0$ and $\alpha\geq 1$.
Assume $2\leq k<|V|$ and $\rho_G(k)>0$.
Let $B_1,\dots,B_k$ be $k$ disjoint subsets
of positive volume in $G$ with
$\max_{i\in[k]}\phi_G(B_i)\leq\alpha\rho_G(k)$.
If
\[
\frac{\rho_G(k+1)}{\rho_G(k)}\ge
\frac{18(\alpha+1)(1+\eps)}{\eps},
\]
then Algorithm \ref{alg:eps-approx},
using $O(k(1+\log\alpha))$ minimum $s$-$t$
cut computations, finds $k$  
disjoint nonempty subsets $C_1,\dots,C_k$
with
\[
\max_{i\in[k]}\phi_G(C_i)
\leq(1+\eps)\rho_G(k).
\]
\end{theorem}

By Theorem~\ref{thm:epsapprox}, 
a   conductance gap of
$\Omega(\alpha/\eps)$ suffices to boost
any $\alpha$-approximate solution to a
$(1+\eps)$-approximate solution for
$\rho_G(k)$.
This improvement requires only
$O(k(1+\log\alpha))$ minimum $s$-$t$
cut computations beyond the initial
approximation algorithm.  
Note that there are almost-linear time algorithms for exact
$s$-$t$ minimum cut   for graphs with
polynomially bounded integral capacities
(see, e.g.,~\cite{CKLPGS22,vdBCKLPGSS23}).
Hence, combining Theorems~\ref{thm:main} and~\ref{thm:epsapprox} gives us the following corollary.

\begin{corollary}\label{cor:largegap}
   Let $G=(V,E,w)$ be a weighted graph, $0<\eps\leq 1$ and $\alpha\geq 1$.
Assume $2\leq k<|V|$ and $\rho_G(k)>0$. Assume that $A_1,\dots, A_k$ are $k$ disjoint subsets with $\max_{i\in [k]} \phi_G(A_i)\leq \alpha \rho_G(k)$. If \[\frac{\rho_G(k+1)}{\rho_G(k)}\geq \frac{18(\alpha+1)(1+\eps)}{\eps},\] then in   $O\left(k(1+\log\alpha)m^{1+o(1)}\right)$ time we can obtain
    \begin{itemize}
    \item [\textup{(i)}] $k$ pairwise disjoint subsets $B_1,\dots,B_k$ such that it holds for every $i\in[k]$ that 
    \[
    \phi_G(B_i)\leq
    \frac{17(1+\eps)\rho_G(k)}{16}
    \qquad\text{and}\qquad
    \phi(G\{B_i\})\geq
    \frac{\rho_G(k+1)}{9};
    \]
    \item [\textup{(ii)}] a partition $V=P_1\sqcup\cdots\sqcup P_k$ such that it holds for every $i\in[k]$ that 
    \[
    \phi_G(P_i)\leq4(1+\eps)k\rho_G(k)
    \qquad\text{and}\qquad
    \phi(G\{P_i\})\geq
    \frac{\rho_G(k+1)}{53k}.
    \]
\end{itemize}
\end{corollary}
An important feature of our
expander-construction algorithms is that
they do not require the values of
$\rho_G(k)$ or $\rho_G(k+1)$.
The conductance-gap assumption is used
only in the analysis and need not be
verified by the algorithms.
Moreover, the value of $\rho_G(k+1)/\rho_G(k)$ can be approximated by the eigenvalues of the normalized Laplacian matrix  $\mathcal{L}_G$  of $G$ and the higher-order Cheeger inequality.

We further study the cluster decomposition problem for graphs without the condition of $\rho_G(k+1)/\rho_G(k)$.  Oveis Gharan and Trevisan~\cite[Theorem~1.5]{GT14} presented an algorithm that finds a cluster decomposition with the bi-criteria guarantee.  Specifically, for every $k\geq2$ with
 the $k$-th smallest eigenvalue $\lambda_k(\mathcal{L}_G) $ of $\mathcal{L}_G$ satisfying $\lambda_k(\mathcal{L}_G)>0$, their algorithm finds
a partition
$V=P_1\sqcup\cdots\sqcup P_\ell$, for some $1\leq\ell<k$, satisfying
\[\phi_G(P_i)=O\left(k^6\sqrt{\lambda_{k-1} (\mathcal{L}_G)}\right)\qquad\text{and}\qquad
\phi(G\{P_i\})=\Omega\left(\frac{\lambda_k^2(\mathcal{L}_G)}{k^4}\right) 
\]
for every $i\in[\ell]$. Their algorithm runs in $\TildeO\left(knm^2\cdot  w_{\max}/w_{\min}\right)$ time, where $w_{\max}$ and $w_{\min}$ are the maximum and minimum weights of the edges respectively. For over a decade, this was the only algorithm for cluster decomposition that provided both inner and outer conductance guarantees on the output clusters.



As the third result, we present a 
nearly-linear time algorithm for the problem, and the asymptotic approximation guarantee of our algorithm matches the one of theirs.
To obtain this result, we replace the potentially  long local-search procedures
with   Algorithms~\ref{alg:fair-trimming}
and \ref{alg:fair-attachment}
developed in Theorem~\ref{thm:main}.
Our algorithm thus provides a
nearly-linear time implementation of
the spectral partitioning framework.  Our result is summarized as follows:

\begin{theorem} 
\label{thm:fast-GT}
For every $k\geq 2$ with
$\lambda_k(\mathcal{L}_G)>0$,
suppose that the spectral parameters
$\lambda_{k-1}(\mathcal{L}_G)$ and
$\lambda_k(\mathcal{L}_G)$ are given
to the algorithm. Then, there is a randomized
algorithm (Algorithm \ref{alg:fast-GT}) that runs in  
$
\TildeO(mk^2)$ time
and outputs a partition
$V=P_1\sqcup\cdots\sqcup P_\ell$
for some $1\leq\ell<k$, satisfying
\[
\phi_G(P_i)
=
O\left(
k^6\sqrt{\lambda_{k-1}(\mathcal{L}_G)}
\right)
\qquad\text{and}\qquad
\phi(G\{P_i\})
=
\Omega\left(
\frac{\lambda_k^2(\mathcal{L}_G)}{k^4}
\right)
\]
for every $i\in[\ell]$.
\end{theorem}

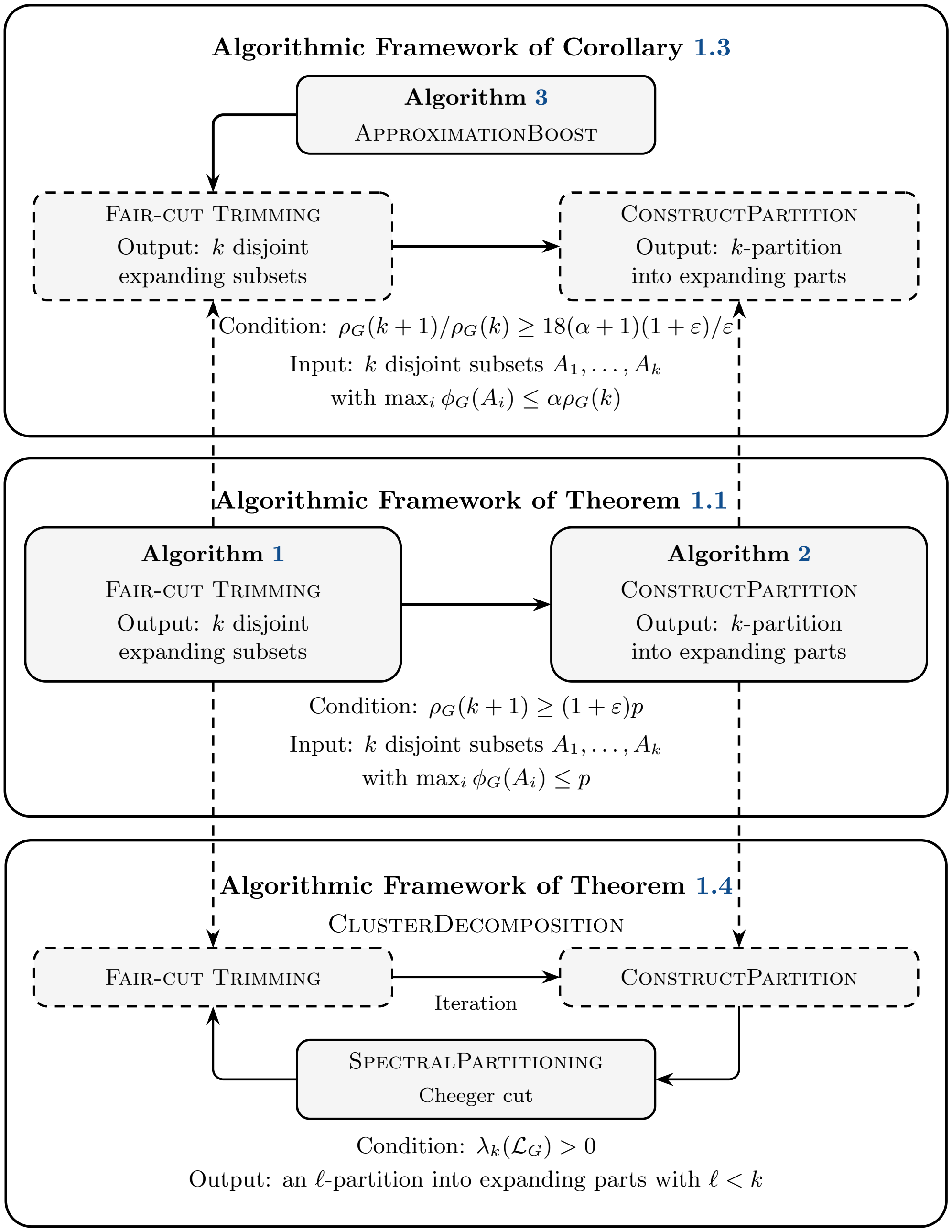
\begin{figure*}[!ht]
\centering
\providecommand{\eps}{\varepsilon}

\resizebox{0.8\textwidth}{!}{%

\begin{tikzpicture}[
  font=\footnotesize,
  text=black,
  card/.style={
    draw=black,
    fill=black!4,
    line width=0.9pt,
    rounded corners=8pt,
    minimum width=5.0cm,
    minimum height=1.15cm,
    text width=4.65cm,
    align=center,
    inner sep=7pt,
    outer sep=0pt
  },
  module/.style={
    draw=black,
    fill=black!4,
    dashed,
    dash pattern=on 4pt off 3pt,
    line width=0.9pt,
    rounded corners=5pt,
    minimum width=4.9cm,
    minimum height=0.8cm,
    text width=4.55cm,
    align=center,
    inner sep=5pt,
    outer sep=0pt
  },
  data/.style={
    -{Stealth[length=2.5mm,width=1.8mm]},
    draw=black,
    line width=1.1pt
  },
  reuse/.style={
    -{Stealth[length=2.5mm,width=1.8mm]},
    draw=black,
    dashed,
    dash pattern=on 4pt off 3pt,
    line width=1pt
  }
]


\path[
  draw=black,
  fill=black!0,
  line width=1pt,
  rounded corners=10pt
] (-6.45,3.20) rectangle (6.45,9.1);

\node[
  font=\small\bfseries,
  align=center
] at (0,8.5) {
  Algorithmic Framework of
  Corollary~\ref{cor:largegap}
};

\node[module,solid] (boost) at (0,7.6) {
  \textbf{Algorithm~\ref{alg:eps-approx}}\\[3pt]
  \textsc{ApproximationBoost}
};

\node[module] (toptrim) at (-3.6,5.8) {
  \textsc{Fair-cut Trimming}\\[2pt]
  Output: $k$ disjoint expanding subsets
};

\node[module] (toppart) at (3.6,5.8) {
  \textsc{ConstructPartition}\\[2pt]
  Output: $k$-partition into expanding parts
};

\draw[data,rounded corners=4pt]
  (boost.west) -| (toptrim.north);

\draw[data]
  (toptrim.east) -- (toppart.west);

\node[ font=\footnotesize,
  align=center] at (0,4.20) {
  Condition:
  $\displaystyle
  \rho_G(k+1)/\rho_G(k)
  \geq
  18(\alpha+1)(1+\eps)/\eps$
  \\[4pt]
  Input:
  $k$ disjoint subsets $A_1,\ldots,A_k$\\[2pt]
  with $\max_i\phi_G(A_i)\leq\alpha\rho_G(k)$
};


\path[
  draw=black,
  fill=black!0,
  line width=1pt,
  rounded corners=10pt
] (-6.45,-2) rectangle (6.45,2.9);

\node[
  font=\small\bfseries,
  align=center
] at (0,2.3) {
  Algorithmic Framework of
  Theorem~\ref{thm:main}
};

\node[
  card,
  minimum height=1.65cm
] (algone) at (-3.6,0.9) {
  \textbf{Algorithm~\ref{alg:fair-trimming}}\\[3pt]
  \textsc{Fair-cut Trimming}\\[2pt]
   Output: $k$ disjoint expanding subsets
};

\node[
  card,
  minimum height=1.65cm
] (algtwo) at (3.6,0.9) {
  \textbf{Algorithm~\ref{alg:fair-attachment}}\\[3pt]
  \textsc{ConstructPartition}\\[2pt]
  Output: $k$-partition into expanding parts
};

\draw[data]
  (algone.east) -- (algtwo.west);

\node[ font=\footnotesize,
  align=center] at (0,-1) {
  Condition: $\rho_G(k+1)\geq(1+\eps)p$\\[4pt]
  Input: $k$ disjoint subsets $A_1,\ldots,A_k$\\[2pt]
  with $\max_i\phi_G(A_i)\leq p$
};


\node[
  font=\small,
  align=center
] (a3title) at (0,-3.2) {
  \textbf{Algorithmic Framework of
  Theorem \ref{thm:fast-GT}}\\[3pt]
  \textsc{ClusterDecomposition}
};

\node[module] (trim) at (-3.6,-4.2) {
  \textsc{Fair-cut Trimming}
};

\node[module] (partition) at (3.6,-4.2) {
  \textsc{ConstructPartition}
};

\node[module,solid] (spectral) at (0,-5.6) {
  \textsc{SpectralPartitioning}\\[2pt]
  {\scriptsize Cheeger cut}
};

\draw[data,line width=0.85pt]
  (trim.east) -- (partition.west);

\draw[data,line width=0.85pt,rounded corners=4pt]
  (partition.south) |- (spectral.east);

\draw[data,line width=0.85pt,rounded corners=4pt]
  (spectral.west) -| (trim.south);

\node[
  font=\scriptsize
] at (0,-4.55) {
  Iteration
};

\node[
  font=\footnotesize,
  align=center
] (a3output) at (0,-6.75) {
  Condition: $\lambda_k(\mathcal{L}_G)>0$\\[2pt]
  Output: an $\ell$-partition into expanding parts with
  $\ell<k$
};

\begin{scope}[on background layer]
  \node[
    draw=black,
    fill=white,
    line width=1pt,
    rounded corners=10pt,
    fit=(a3title)(trim)(partition)(spectral)(a3output),
    inner xsep=11pt,
    inner ysep=10pt
  ] (algthree) {};
\end{scope}


\draw[reuse]
  (algone.north) -- (toptrim.south);

\draw[reuse]
  (algtwo.north) -- (toppart.south);


\draw[reuse]
  (algone.south) -- (trim.north);

\draw[reuse]
  (algtwo.south) -- (partition.north);
\end{tikzpicture}
}

\caption{Overview of our algorithmic framework. The dashed arrows and boxes indicate reuse of algorithmic procedures.
}
\label{fig:algorithm-framework}
\end{figure*}

Figure~\ref{fig:algorithm-framework} illustrates the connections of our developed four algorithms.

Finally, it is worth comparing the approximation guarantee of 
Corollary~\ref{cor:largegap}(ii) and
Theorem~\ref{thm:fast-GT}.
While Theorem~\ref{thm:fast-GT}
requires no conditions on $\rho_G(k+1)$ and $\rho_G(k)$,
Corollary~\ref{cor:largegap}(ii)
provides asymptotically stronger inner
and outer conductance guarantees under
a sufficiently large conductance gap,
assuming that an approximation for $\rho_G(k)$ is given.
More precisely, compared with Theorem~\ref{thm:fast-GT}
applied with parameter $k+1$,
Corollary~\ref{cor:largegap}(ii)
not only produces a partition into
exactly $k$ clusters, but also provides
improved guarantees on the conductance
of each cluster.
For example, suppose that
$2\leq k <|V|$, $\rho_G(k)>0$,
and an $\alpha$-approximation for 
$\rho_G(k)$ is given, where $\alpha\geq1$.
Setting $\eps=1$ in
Corollary~\ref{cor:largegap},
if the conductance gap satisfies
$\rho_G(k+1)/\rho_G(k)>36(\alpha+1)$,
then our algorithms find a partition
$V=P_1\sqcup\cdots\sqcup P_k$
such that for every $i\in[k]$,
\[
\phi_G(P_i)
\leq 8k\rho_G(k)
\leq O\left(
k^3\sqrt{\lambda_k(\mathcal{L}_G)}
\right)
\qquad\text{and}\qquad
\phi(G\{P_i\})
\geq
\frac{\rho_G(k+1)}{53k}
\geq\Omega\left(
\frac{\lambda_{k+1}(\mathcal{L}_G)}{k}
\right),
\]
where the last inequalities
follow from the higher-order Cheeger
inequality~\cite{LOT14}.
In contrast,
Theorem~\ref{thm:fast-GT}, when applied
with parameter $k+1$, only guarantees
a partition with  some $\ell\leq k$ clusters,
with outer conductance
$O\left((k+1)^6\sqrt{\lambda_k(\mathcal{L}_G)}\right)$
and inner conductance
$\Omega\left(\lambda_{k+1}^2(\mathcal{L}_G)/(k+1)^4\right)$.
Moreover, the value of 
$\alpha$ can be instantiated using any
approximation algorithm for
$k$-way conductance that returns exactly
$k$ disjoint subsets.
In particular, when $k=2$,  the classical
algorithm of Arora, Rao, and
Vazirani~\cite{ARV09} gives an
$O\left(\sqrt{\log n}\right)$-approximation.
Consequently, if
$\rho_G(3)/\rho_G(2)>C\sqrt{\log n}$
for a sufficiently large constant $C$,
our algorithms find a partition
$V=P_1\sqcup P_2$ such that it holds 
for every $i\in[2]$ that 
\[
\phi_G(P_i)
\leq16\rho_G(2)
\leq O\left(
\sqrt{\lambda_2(\mathcal{L}_G)}
\right)
\qquad\text{and}\qquad
\phi(G\{P_i\})
\geq
\frac{\rho_G(3)}{106}
\geq\Omega\left(
\lambda_3(\mathcal{L}_G)
\right).
\]

\subsection{Related Work}

Our work relates to a sequence of results on \emph{expander decomposition}, which is a fundamental tool in the design of fast graph algorithms    with applications
to graph sparsification, Laplacian
system solvers, flow algorithms, and
dynamic graph algorithms~\cite{ST04,ST11,SW19}. Given an input graph $G$ and parameter $\varphi$,  a $\varphi$-expander
decomposition partitions $V$ into
subsets $P_1,\ldots,P_\ell$ such that
each induced subgraph $G[P_i]$ has inner
conductance at least $\varphi$,
while the total weight of inter-cluster 
edges is small. Spielman and Teng~\cite{ST04}
developed a nearly-linear time algorithm
for a weaker form of expander decomposition.
For unweighted graphs, a simple recursive
sparse-cut argument shows that every graph
admits a $\varphi$-expander decomposition
with at most $O(\varphi m\log n)$
crossing edges, and this bound is
asymptotically tight in general~\cite{ST11}.
Saranurak and Wang~\cite{SW19}
subsequently obtained a nearly-linear-time
algorithm producing clusters that each
induce an expander, with
$O(\varphi m\log^3 n)$ crossing edges.
Agassy, Dorfman, and Kaplan~\cite{ADK23}
improved this bound to
$O(\varphi m\log^2 n)$ while retaining
a running time of $\widetilde O(m/\varphi)$.
More recently, Bansal, Jambulapati,
and Saranurak~\cite{BJS26}
obtained a polynomial-time algorithm
with only $O(\varphi m\log^{1+o(1)}n)$
crossing edges, nearly matching the
existential bound. Their result further guarantees the
stronger notion of flow expansion.

Our work also relates to the \emph{$k$-way conductance
minimization}, which seeks $k$  
disjoint subsets with low outer
conductance. More precisely, the objective is to find $k$  
disjoint subsets $A_1,\dots,A_k$ with $\max_{1\leq i\leq k}\phi_G(A_i)$ as close to $\rho_G(k)$ as possible.
For $k=2$, this problem reduces to the
classical graph conductance problem, since
$\rho_G(2)=\phi(G)$.
Arora, Rao, and Vazirani~\cite{ARV09}
developed a polynomial-time
$O(\sqrt{\log n})$-approximation algorithm
for this problem.
For general $k$, the study of $k$-way
conductance minimization is   connected to
higher-order Cheeger inequalities and
spectral graph partitioning.
Lee, Oveis Gharan, and Trevisan~\cite{LOT14}
established the following constructive
higher-order Cheeger inequality:
\[
\frac{\lambda_k\ (\mathcal{L}_G)}{2}
\leq \rho_G(k)
\leq O\left(k^2\sqrt{\lambda_k (\mathcal{L}_G) }\right).
\]
A complementary result by Louis, Raghavendra,
Tetali, and Vempala~\cite{LRTV12} achieved
the improved conductance bound
$O(\sqrt{\lambda_k (\mathcal{L}_G) \log k})$, at the expense
of guaranteeing only $\Omega(k)$ rather than
$k$ pairwise disjoint subsets.
Both results are algorithmic and run in
polynomial time.
For the more restrictive problem of
partitioning the entire vertex set,
Louis and Makarychev~\cite{LM14} developed
a polynomial-time bicriteria approximation
algorithm that produces at least
$(1-\eta)k$ parts, each with conductance
at most $O_\eta(\sqrt{\log n\log k})$
times the optimal $k$-partitioning value $\mathsf{OPT_k}$, i.e., 
\[\mathsf{OPT_k}\triangleq \min_{\substack{
V=P_1\sqcup\cdots\sqcup P_k\\
\vol_G(P_i)>0\;\forall i\in[k]
}}
\max_{1\leq i\leq k}\phi_G(P_i).\]

These two lines of research address
complementary aspects of graph clustering,
but neither alone guarantees that clusters
are simultaneously well separated and
internally expanding.
Minimizing $k$-way conductance produces
$k$ disjoint subsets with small outer
conductance, but these subsets need not
be internally well connected. In contrast, expander decomposition
guarantees high inner conductance, but
does not prescribe the number of clusters
or necessarily ensure low outer
conductance for each individual cluster.  Table~\ref{tab:comparison} summarizes
the guarantees of the related results
discussed above, and compares them with
our algorithms in terms of output type,
inner and outer conductance, crossing
edges, and running time.

\begin{table*}[h]
\centering
\caption{Comparison between our results and the related ones.}
\vspace{0.5em}
\label{tab:comparison}
\footnotesize
\setlength{\tabcolsep}{5pt}
\renewcommand{\arraystretch}{1.2}

\resizebox{\textwidth}{!}{
\begin{tabular}{@{}lccccc@{}}
\toprule
\textbf{Reference} & \textbf{Type of Output}
& \textbf{Inner conductance}
& \textbf{Outer conductance}
& \textbf{\# Crossing edges}
& \textbf{Running time} \\
\midrule

\multicolumn{6}{@{}l}{\emph{Results with only the inner conductance guarantee}}\\[2pt]
\cmidrule{1-1}

\cite{ST11}
& \multirow{4}{*}{\shortstack{partition\\ (with an unknown\\ number of parts)}}
& \multirow{4}{*}{$\geq\varphi$}
& \multirow{4}{*}{$-$}
& $O(\varphi m\log n)$
& Existential Result \\

\cite{SW19}
&
&
&
& $O(\varphi m\log^3 n)$
& \multirow{2}{*}{$\widetilde O(m/\varphi)$}\\

\cite{ADK23}
&
&
&
& $O(\varphi m\log^2 n)$
& \\

\cite{BJS26}
&
&
&
& $O(\varphi m\log^{1+o(1)} n)$
& $\operatorname{poly}(n)$ \\

\midrule
\multicolumn{6}{@{}l}{\emph{Results with only the outer conductance guarantee}}\\[2pt]
\cmidrule{1-1}

\cite{LRTV12}
& $\Omega(k)$ disjoint subsets
& \multirow{3}{*}{$-$}
& $O\left(\sqrt{\lambda_k (\mathcal{L}_G) \log k}\right)$
& \multirow{3}{*}{$-$}
& \multirow{2}{*}{$\operatorname{poly}(n)$} \\

\cite{LOT14}
& $k$ disjoint subsets
&
& $O\left(k^2\sqrt{\lambda_k (\mathcal{L}_G)} \right)$
&
& \\

\cite{LM14}
& $\ell$-partition, $\ell\geq(1-\eta)k$
&
& $O_\eta(\sqrt{\log n\log k})\,
  \mathrm{OPT}_k$
&
& $\operatorname{poly}_{\eta}(n)$\\

\midrule
\multicolumn{6}{@{}l}{\emph{Results with both inner and outer conductance guarantees}}\\[2pt]
\multicolumn{6}{@{}l}{Condition:
\emph{$\rho_G(k+1)/\rho_G(k)>1+\eps$ and given $k$ disjoint subsets $A_1,\dots, A_k$ with $\max_{i\in [k]}\phi_G(A_i)=\rho_G(k)$}}\\[2pt]
\cmidrule{1-1}

\cite{GT14}
& \multirow{2}{*}{$k$ disjoint subsets}
& \multirow{2}{*}{$\Omega(\eps\rho_G(k+1))$}
& \multirow{2}{*}{$O(\rho_G(k))$}
& \multirow{2}{*}{$-$}
& Existential Result \\

\textbf{Theorem \ref{thm:main} (i)}
&
&
&
&
& $\bm{\widetilde O(m/\eps)}$ \\
\cmidrule{2-6}

\cite{GT14}
& \multirow{2}{*}{$k$-partition}
& \multirow{2}{*}{$\Omega(\eps\rho_G(k+1)/k)$}
& \multirow{2}{*}{$O(k\rho_G(k))$}
& \multirow{2}{*}{$O(k\rho_G(k)m)$}
& Existential Result \\

\textbf{Theorem \ref{thm:main} (ii)}
&
&
&
&
& $\bm{\widetilde O(m(\eps^{-1}+\log^2 k))}$ \\
\midrule
\multicolumn{6}{@{}l}{Condition: \emph{$\rho_G(k+1)/\rho_G(k)>18(\alpha+1)(1+\eps)/\eps$ and given $k$ disjoint subsets $A_1,\dots, A_k$ with $\max_{i\in [k]}\phi_G(A_i)\leq \alpha\rho_G(k)$}}\\[2pt]
\cmidrule{1-1}

\multirow{2}{*}{\textbf{Corollary \ref{cor:largegap}}}
& $k$ disjoint subsets
& $\Omega(\rho_G(k+1))$
& $O(\rho_G(k))$
& $-$
&  \multirow{2}{*}{\bm{$O\left(k(1+\log\alpha)m^{1+o(1)}\right)$}}\\

& $k$-partition
& $\Omega(\rho_G(k+1)/k)$
& $O(k\rho_G(k))$
& $O(k\rho_G(k)m)$
& \\

\midrule
\multicolumn{6}{@{}l}{Condition: \emph{$\lambda_k(\mathcal{L}_G)>0$}}\\[2pt]
\cmidrule{1-1}

\cite{GT14}
& \multirow{2}{*}{$\ell$-partition, $\ell< k$}
& \multirow{2}{*}{$\Omega\left(\lambda_k^2(\mathcal{L}_G)/k^4\right)$}
& \multirow{2}{*}{$O\left(k^6\sqrt{\lambda_{k-1}(\mathcal{L}_G)}\right)$}
& \multirow{2}{*}{$O\left(k^6\sqrt{\lambda_{k-1}(\mathcal{L}_G)}m\right)$}
& $\operatorname{poly}(n)$\\

\textbf{Theorem \ref{thm:fast-GT}}
&
&
&
&
& $\bm{\widetilde O(mk^2)}$ \\

\bottomrule
\end{tabular}
}
\vspace{0.5em}
\begin{minipage}{0.98\textwidth}
\scriptsize
\end{minipage}
\end{table*}

\section{Preliminaries\label{sec:pre}}

\subsection{Notation}\label{sec:notation}

Let $G=(V,E,w)$ be an undirected and  weighted graph (loops allowed) with a positive weight function $w:E\to\mathbb{R}_{> 0}$. For a vertex $v\in V$, its degree   is defined by
$
\deg(v)\triangleq \sum_{u\sim v} w(u,v)$.
For subsets $S,T\subseteq V$, let
$
w(S,T) \triangleq \sum_{\substack{uv\in E\\u\in S,\;v\in T}} w(u,v)$.
The volume of a subset $S\subseteq V$ is defined by  
$\vol(S) \triangleq \sum_{v\in S}\deg(v)$.



Our analysis is based on the spectral properties of graphs, and we list the basics of spectral graph theory. For a graph $G=(V,E,w)$ with $n$ vertices, let $D_G \in \mathbb{R}^{n \times n}$ be the degree matrix with $D_G(u,u) = \deg(u)$, and $ A_G \in \mathbb{R}^{n \times n}$ the adjacency matrix with $A_G(u,v) = w(u,v)$. 
The normalized Laplacian of $G$ is
$
\mathcal{L}_G \triangleq I -N_G,
$
where  $I$ is the identity matrix and $N_G\triangleq D_G^{-1/2} A_G D_G^{-1/2}$. By definition $\mathcal{L}_G$ is symmetric and positive semidefinite, with real eigenvalues 
\[
0 = \lambda_1 (\mathcal{L}_G)  \le \cdots \le \lambda_n (\mathcal{L}_G)  \le 2,
\]
and an associated orthonormal basis of eigenvectors
$f_1, \dots, f_n \in \mathbb{R}^n$.
For any $f:V(G)\to\mathbb{R}$ satisfying $f\not\equiv 0$, the Rayleigh quotient of $f$ is defined by
\[
\RQ_G(f)\triangleq \frac{\sum_{uv\in E,\,u\ne v} w(u,v)\,(f(u)-f(v))^2}{\sum_{v\in V}\deg(v)f(v)^2}.
\]
Recall that, by the Courant--Fischer formulation, it holds for every $1\le k\le n$ that 
\begin{equation}\label{eq:CF}
\lambda_k(\mathcal{L}_G)=\min_{\substack{U\subseteq \mathbb{R}^{n}\\ \dim U=k}}\ \max_{0\ne f\in U}\RQ_G(f).
\end{equation}


\subsection{Useful Results}

In this subsection we present a sequence of results that will be used in our analysis, and the  proofs of Lemmas \ref{lem:disjoint-rayleigh}--\ref{lem:fixed11}  can be found in Section \ref{sec:pre_app}. We first state the Cheeger inequality for graphs and the approximation guarantee of the spectral partitioning algorithm.

\begin{lemma}[The Cheeger inequality]
\label{lem:cheeger-cut}
Let $G$ be a weighted graph, possibly with self-loops, with positive weighted degrees and $|V(G)|\geq2$. Then, it holds that 
\[
\frac{\lambda_2(\mathcal{L}_G)}{2}
\leq
\phi(G)
\leq
\sqrt{2\lambda_2(\mathcal{L}_G)}.
\]
Moreover, given the   eigenvector corresponding to $\lambda_2(\mathcal{L}_G)$ as input, there is a nearly-linear time  algorithm called \textsc{SpectralPartitioning} that  returns a set $S$ with
$\vol(S)\leq\vol(G)/2$ and
\[
\phi_G(S)\leq\sqrt{2\lambda_2(\mathcal{L}_G)}.
\]
\end{lemma}

 The following lemma was shown in \cite{KLLGT13}. We present  a complete proof in Section~\ref{apx:sec2} to emphasize that the result remains valid  for weighted graphs with self-loops.

\begin{lemma}[\cite{KLLGT13}]\label{lem:disjoint-rayleigh}
Let $G$ be a weighted graph, possibly with self-loops, with positive weighted degrees, and let $f_1,\dots,f_k$ be nonzero functions on $V(G)$ with pairwise disjoint supports. Then, it holds that 
\[
\lambda_k(\mathcal{L}_G)
\leq
2\max_{1\leq i\leq k}\RQ_G(f_i).
\]
\end{lemma}

The next result first appeared in \cite{LOT14} in the loopless setting. We present a complete proof in  Section~\ref{apx:sec2} to show that the result remains valid for weighted graphs with self-loops.  Our proof is based on a simple reduction to the loopless case.

\begin{lemma} [\cite{LOT14}]\label{lem:LOT}
There is a universal constant $c_0>0$ such that, for every weighted
graph $G$, possibly with self-loops and with positive weighted degrees,
and every $1\leq k\leq |V(G)|$, there exist $k$ nonzero functions
$g_1,\ldots,g_k:V(G)\mapsto\mathbb R$ with pairwise disjoint supports such
that
\[
\RQ_G(g_i)\leq c_0 k^6\lambda_k(\mathcal{L}_G)
\qquad\text{for every }i\in[k].
\]
\end{lemma}

Finally, the following lemma is a weaker version of Lemma~1.1 in \cite{GT14}, and its proof is essentially the same as the one in \cite{GT14}.   In Section~\ref{sec:cex} we present a counterexample showing that the original statement of Lemma~1.1 in \cite{GT14} is incorrect. 
 
\begin{lemma}\label{lem:fixed11}
There is a universal constant $c_0>0$ such that for any $|V|\ge k\ge 2$ and any partition $V=P_1\sqcup\cdots\sqcup P_{\ell}$ for some $\ell\leq k-1$, we have 
\[
\min_{1\le i\le \ell}\lambda_2\left(\mathcal{L}_{G\{P_i\}}\right)
\le
2c_0k^6\lambda_k(\mathcal{L}_G).
\]
\end{lemma}

\section{Cluster Decomposition from Disjoint Subsets}\label{sec:main}

 In this section, we prove that, given $k$ disjoint vertex sets $A_1,\ldots, A_k$ satisfying $
\phi_G(A_i)\le p
$ for every $i\in[k]$ as input and assuming
 $\rho_G(k+1)\ge (1+\eps)p$, there is a nearly-linear time algorithm that finds a partition $P_1,\ldots, P_k$ such that $\phi_G(P_i)\le 4kp$ and $\phi(G\{P_i\})\ge
\frac{\eps\,\rho_G(k+1)}{53k}$. This represents the \emph{first} nearly-linear time algorithm that partitions an input graph into expanders with bounded inner and outer conductance.

The design of our algorithm is based on the notion of \emph{fair cuts} introduced by Li, Nanongkai, Panigrahi, and Saranurak \cite{LNPS23}. Given any undirected and capacitated graph $G$ with terminals $s,t$, an $\alpha$-fair cut $(S,T)$ is an $s,t$-cut admitting a
feasible signed $s,t$-flow $f$ such that, for every cut edge
$(u,v)$ oriented from $S$ to $T$,
\[
f(u,v)\ge \frac{c(u,v)}{\alpha}.
\]
In particular, every $\alpha$-fair $s,t$-cut is an $\alpha$-approximation to the minimum $s,t$-cut; i.e.,  the minimum-capacity cut separating $s$ and $t$ with its cut value denoted by $\operatorname{mincut}(s,t)$. It is known that a $(1+\eta)$-fair $s,t$-cut can be computed in $\TildeO(m/\eta)$ time~\cite{LiLi24}, and this algorithm will be used as a black box in our analysis. The following simple proposition will be used in our analysis.

\begin{proposition}\label{prop:fair-approx}
Every $\alpha$-fair $s,t$-cut is an $\alpha$-approximation to the minimum $s,t$-cut.  
\end{proposition}

\begin{proof}
Suppose the $\alpha$-fair $s,t$-cut has capacity $C$. By the definition of fairness, its witnessing $s,t$-flow has value at least $C/\alpha$. By the max-flow min-cut theorem,
$
C/\alpha\le \operatorname{mincut}(s,t)$,
 and this proves the proposition.
\end{proof}


\subsection{Constructing a Backbone}

We first show that,  given $k$ disjoint vertex sets $A_1,\ldots, A_k$ satisfying $
\phi_G(A_i)\le p
$ for every $i\in[k]$ as input and assuming
 $\rho_G(k+1)\ge (1+\eps)p$, there is a nearly-linear time algorithm that finds $B_1,\ldots, B_k$ with $B_i\subseteq A_i$, and every $B_i$ has high inner conductance and the outer conductance linear with $\phi_G(A_i)$. As our algorithm is mainly based on computing an approximate fair cut, we call the algorithm \textsc{Fair-Cut Trimming},  and its input is an arbitrary set $A\in\{A_i: i\in[k]\}$. See Algorithm~\ref{alg:fair-trimming} for a formal description of \textsc{Fair-Cut Trimming}. Notice that, as  a $(1+\eta)$-fair $s,t$-cut can be computed in  $\TildeO(m/\eta)$ time~\cite{LiLi24} and $\alpha=1+\eps/16$, Algorithm~\ref{alg:fair-trimming} runs in $\TildeO(m/\eps)$ time.

\begin{algorithm}[ht]
\caption{\textsc{Fair-Cut Trimming}}
\label{alg:fair-trimming}
\begin{algorithmic}[1]
\Require A positive-volume subset $A\subseteq V$ with $\phi_G(A)\leq p$
\Ensure A positive-volume backbone $B\subseteq A$

\State Set
\[
\alpha \gets 1+\frac{\eps}{16},
\qquad
\beta \gets \left(1+\frac{\eps}{10}\right)p.
\]

\State Construct an undirected auxiliary graph $H_A$ with
\[
V(H_A)=A\cup\{s,t\}
\]
as follows:
\For{each edge $\{u,v\} \in E$ with $u,v\in A$}
    \State Add the edge $\{u,v\}$ to $H_A$ with capacity $w(u,v)$.
\EndFor
\For{each vertex $v\in A$}
    \State Add an edge $\{s,v\}$ with capacity $w(v,V\setminus A)$.
    \State Add an edge $\{v,t\}$ with capacity $\beta \deg(v)$.
\EndFor

\State Compute an $\alpha$-fair $s,t$-cut $(S,T)$ of $H_A$, with
$s\in S$ and $t\in T$.

\State $B\gets T\setminus\{t\}$.

\State \Return $B$.
\end{algorithmic}
\end{algorithm}


Now we analyze the property of the algorithm's output. Note that for any $s,t$-cut $(S,T)$ of $H_A$, with $s\in S$, $t\in T$, and $B=T\setminus\{t\}$, its capacity is
\begin{equation}\label{eq:stcapcity}
c_{s,t}(B)
\triangleq
w(B,V\setminus B)
+
\beta\vol(A\setminus B).
\end{equation}
Here, the first term measures the boundary of the retained set $B$ in the original graph, while the second term penalizes the volume discarded from $A$. Thus, minimizing the $s,t$-cut capacity balances preserving a low-conductance subset against removing too much of $A$. This trade-off is the main intuition behind the construction of $H_A$.

To establish the properties of the backbone $B$ returned by Algorithm~\ref{alg:fair-trimming}, we first show that $B$ is   positive-volume and   its conductance increases by at most a factor of $\alpha$ compared with that of $A$.

\begin{lemma}\label{lem:backbone-low}
Let $B$ be the backbone returned by Algorithm~\ref{alg:fair-trimming}. Then, $\vol(B)>0$ and
\[
\phi_G(B)\leq \alpha p
=
\left(1+\frac{\eps}{16}\right)p.
\]
\end{lemma}

\begin{proof}
Since $\phi_G(A)\leq p$, the cut $(\{s\},A\cup\{t\})$ in $H_A$ has capacity $w(A,V\setminus A)\leq p\vol(A)$ and  
\[
\operatorname{mincut}(s,t)\leq p\vol(A).
\]
By Proposition~\ref{prop:fair-approx}, the $\alpha$-fair cut returned by Algorithm~\ref{alg:fair-trimming} is an $\alpha$-approximation to the minimum $s,t$-cut. Therefore, 
\begin{equation}\label{eq:l11}
c_{s,t}(B)
\leq
\alpha\,\operatorname{mincut}(s,t)
\leq
\alpha p\vol(A).
\end{equation}

We first show that $B$ has positive volume. If $\vol(B)=0$, then  \eqref{eq:stcapcity} implies that  $c_{s,t}(B)=\beta\vol(A)$. Since $\beta>\alpha p$ and $\vol(A)>0$, we have
\[
c_{s,t}(B)>\alpha p\vol(A),
\]
contradicting \eqref{eq:l11}. Hence,  $\vol(B)>0$.

Next, we   bound the conductance of $B$. By \eqref{eq:l11} and the definition of $c_{s,t}(B)$,
\[
w(B,V\setminus B)
\leq
\alpha p\vol(A)-\beta\vol(A\setminus B).
\]
Using $\vol(A)=\vol(B)+\vol(A\setminus B)$, we obtain

\begin{equation*}
w(B,V\setminus B)
\leq
\alpha p\vol(B)
+
(\alpha p-\beta)\vol(A\setminus B)
\leq
\alpha p\vol(B),
\end{equation*}
where the last inequality follows from $\beta\geq\alpha p$. Therefore,
\[
\phi_G(B)
=
\frac{w(B,V\setminus B)}{\vol(B)}
\leq
\alpha p,
\]
which proves the statement.
\end{proof}

\begin{lemma}\label{lem:subset-flow}
Let $B$ be the backbone returned by Algorithm~\ref{alg:fair-trimming}. Then, it holds  for every $U\subseteq B$ that 
\[
w(U,V\setminus B)
\leq
\alpha\left(w(U,B\setminus U)+\beta\vol(U)\right).
\]
\end{lemma}

\begin{proof}
Let $f$ be the feasible signed $s,t$-flow witnessing the $\alpha$-fairness of the cut. The cut edges from the source side into $U$ have total capacity $w(U,V\setminus B)$. Indeed, these edges consist of the original edges between $A\setminus B$ and $U$, together with the source edges $\{s,v\}$ for every $v\in U$, whose total capacity is $w(U,V\setminus A)$. 
By the property of  $\alpha$-fairness, each such cut edge carries at least a $1/\alpha$ fraction of its capacity from the source side toward $U$. Hence, the total flow entering $U$ from the source side is at least
\[
\frac{w(U,V\setminus B)}{\alpha}.
\]
Since $U$ contains neither $s$ nor $t$, flow conservation implies that this incoming flow   leaves $U$. Moreover, by the definition of a fair-cut, all cut edges between $T=B\cup\{t\}$ and $S= V(H_A) \setminus T$ are oriented from $S$ to $T$. Therefore, the flow entering $U$ must leave $U$ through either edges between $U$ and $B\setminus U$ or edges $\{v,t\}$ with $v\in U$; the total capacity of these edges is
\[
w(U,B\setminus U)+\beta\vol(U).
\]
Hence,
\[
\frac{w(U,V\setminus B)}{\alpha}
\leq
w(U,B\setminus U)+\beta\vol(U),
\]
and this proves the lemma.
\end{proof}

Next, we prove that  the backbone $B$ returned by Algorithm~\ref{alg:fair-trimming} has strong relative expansion: for every positive-volume bipartition of $B$, the boundary between the two sides is large compared with the boundary leaving $B$. Following \cite{GT14}, for $U\subseteq B$ with
$
0<\vol(U)<\vol(B)$, we 
define the \emph{relative expansion} of $U$ in $B$ by
\[
\varphi(U,B)
\triangleq 
\frac{w(U,B\setminus U)}
{\frac{\vol(B\setminus U)}{\vol(B)}w(U,V\setminus B)}.
\]
If $w(U,V\setminus B)=0$, we use the convention
$\varphi(U,B)=+\infty$.

\begin{lemma}\label{lem:one-shot}
Let $B$ be the backbone returned by Algorithm~\ref{alg:fair-trimming}. Then, for every $U\subseteq B$ satisfying $0<\vol(U)<\vol(B)$ and $T=B\setminus U$, we have
\[
\varphi(U,B)\geq \frac{\eps}{4}.
\]
Equivalently, 
\[
w(U,T)
\geq
\frac{\eps}{4}
\frac{\vol(T)}{\vol(B)}
w(U,V\setminus B).
\]
\end{lemma}

To prove Lemma~\ref{lem:one-shot}, we set 
\[
\sigma\triangleq\frac{\rho_G(k+1)-\alpha\beta}{1+\alpha},
\]
and first prove the following claim.
\begin{claim}\label{cl:sigma}
For $T=B\setminus U$, we have 
\[w(U,T)\geq\sigma\min\{\vol(U),\vol(T)\}\qquad and \qquad
\sigma\geq \frac{2}{5}\eps p.\]
\end{claim}

\begin{proof}
By Lemma~\ref{lem:subset-flow}, $w(U,V\setminus B)\leq \alpha\left(w(U,T)+\beta\vol(U)\right)$. Since
$w(U,V\setminus U)=w(U,T)+w(U,V\setminus B)$, we have
\begin{equation}\label{eq:ut1}
(1+\alpha)w(U,T)
\geq
w(U,V\setminus U)-\alpha\beta\vol(U)
=
\bigl(\phi_G(U)-\alpha\beta\bigr)\vol(U).
\end{equation}
Applying Lemma~\ref{lem:subset-flow} to $T$ similarly gives
\begin{equation}\label{eq:ut2}
(1+\alpha)w(U,T)
\geq
w(T,V\setminus T)-\alpha\beta\vol(T)
=
\bigl(\phi_G(T)-\alpha\beta\bigr)\vol(T).
\end{equation}
Suppose that $B\subseteq A_i$. The $k+1$ sets $A_1,\ldots,A_{i-1},U,T,A_{i+1},\ldots,A_k$ are pairwise disjoint and have positive volume. Moreover, it holds for every
$j\neq i$ that 
\[
\phi_G(A_j)\leq p<(1+\eps)p\leq \rho_G(k+1),
\]
where the strict inequality follows from $p>0$ and $\eps>0$.
Therefore, by the definition of $\rho_G(k+1)$,
\[
\max\{\phi_G(U),\phi_G(T)\}\geq \rho_G(k+1).
\]
Combining this with \eqref{eq:ut1} and \eqref{eq:ut2} yields
\[
w(U,T)
\geq
\frac{\rho_G(k+1)-\alpha\beta}{1+\alpha}
\min\{\vol(U),\vol(T)\}
=
\sigma\min\{\vol(U),\vol(T)\}.
\]

It remains to lower-bound $\sigma$. Since $\rho_G(k+1)\geq(1+\eps)p$,
we have
\[
\rho_G(k+1)-\alpha\beta\geq(1+\eps)p-\left(1+\frac{\eps}{16}\right)
\left(1+\frac{\eps}{10}\right)p=\left(\frac{67}{80}\eps-\frac{\eps^2}{160}\right)p\geq\frac{133}{160}\eps p,
\]
where the last inequality uses $\eps\leq1$. In addition, as $1+\alpha
=2+\eps/16\leq 33/16$,
\[
\sigma
\geq
\frac{16}{33}\cdot\frac{133}{160}\eps p
=
\frac{133}{330}\eps p
>
\frac{2}{5}\eps p.
\]
This proves the claim.
\end{proof}

\begin{proof}[Proof of Lemma~\ref{lem:one-shot}]
 By Lemma~\ref{lem:subset-flow},
\[
\varphi(U,B)
=
\frac{w(U,T)}
{\frac{\vol(T)}{\vol(B)}w(U,V\setminus B)}
\geq
\frac{w(U,T)\vol(B)}
{\alpha\vol(T)\bigl(w(U,T)+\beta\vol(U)\bigr)}.
\]
The analysis is based on a case distinction.
\begin{itemize}
\item
If $\vol(U)\leq\vol(T)$, then by Claim~\ref{cl:sigma}, $w(U,T)\geq\sigma\vol(U)$ and therefore
\[
\varphi(U,B)
\geq
\frac{\vol(B)}
{\alpha\vol(T)
\left(1+\frac{\beta\vol(U)}{w(U,T)}\right)}
\geq
\frac{\vol(B)}
{\alpha\vol(T)(1+\beta/\sigma)}
\geq
\frac{1}{\alpha(1+\beta/\sigma)}.
\]
\item 
If \(\vol(U)>\vol(T)\), then by Claim \ref{cl:sigma} $\frac{w(U,T)}{\vol(T)}\ge\sigma$, and hence
\[
\varphi(U,B)
\ge
\frac{
\frac{w(U,T)}{\vol(T)}
\left(1+\frac{\vol(U)}{\vol(T)}\right)}
{\alpha\left(
\frac{w(U,T)}{\vol(T)}
+\beta\frac{\vol(U)}{\vol(T)}
\right)}\
\ge
\frac{
\frac{w(U,T)}{\vol(T)}
}{
\alpha\left(
\frac{w(U,T)}{\vol(T)}+\beta
\right)}
\ge
\frac{\sigma}{\alpha(\sigma+\beta)}=
\frac{1}{\alpha(1+\beta/\sigma)}.
\]
\end{itemize}
Combining the two cases above, we have 
\begin{equation}\label{eq:phiUB}
\varphi(U,B)\geq\frac{1}{\alpha(1+\beta/\sigma)}.
\end{equation}
By Claim~\ref{cl:sigma}, we have  $\sigma\geq 2\eps p/5$. Combining it with $\beta=\left(1+\eps/10\right)p\leq 11p/10$ gives
\[
\frac{\beta}{\sigma}
\leq
\frac{11p}{10}\cdot\frac{5}{2\eps p}
=
\frac{11}{4\eps}.
\]
Since $\eps\leq1$ and $\alpha\leq17/16$, it follows from \eqref{eq:phiUB} that
\[
\frac{1}{\varphi(U,B)}\leq \alpha\left(1+\frac{\beta}{\sigma}\right)
\leq
\frac{17}{16}
\left(1+\frac{11}{4\eps}\right)
\leq
\frac{17}{16}\cdot\frac{15}{4\eps}
=
\frac{255}{64\eps}
<
\frac{4}{\eps},
\]
completing the proof.
\end{proof}

The following result is a useful corollary of Lemma \ref{lem:one-shot} whose analysis essentially follows \cite{GT14}.

\begin{corollary}\label{cor:relative}
Let $B$ be the backbone returned by Algorithm~\ref{alg:fair-trimming}. Then, for every $U\subseteq B$ satisfying $0<\vol(U)\leq\vol(B)/2$ and $T=B\setminus U$, we have
\[
\frac{w(U,T)}{\vol(U)}
\ge
\frac{\varepsilon}{9}
\max\{\phi_G(U),\phi_G(T)\}.
\]
\end{corollary}

\begin{proof}
Since \(\vol(U)\le \vol(T)\), then it holds by Lemma~\ref{lem:one-shot} that $w(U,T)\ge\frac{\varepsilon}{8}w(U,V \setminus B)$, 
which implies 
\[
\phi_G(U)
=
\frac{w(U,T)+w(U,V \setminus B)}{\vol(U)}
\le
\frac{8+\eps}{\varepsilon}\frac{w(U,T)}{\vol(U)}\leq \frac{9}{\varepsilon}\frac{w(U,T)}{\vol(U)}.
\]
Applying Lemma \ref{lem:one-shot} to \(T\), we have
\[
w(T,V\setminus B)
\le
\frac{4w(U,T)(\vol(U)+\vol(T))}{\varepsilon \vol(U)}.
\]
Since \(\vol(T)\ge \vol(U)\), we have 
\begin{eqnarray*}
\phi_G(T)&\leq &\frac{w(U,T)}{\vol(T)}+\frac{4w(U,T)(\vol(U)+\vol(T))}{\varepsilon \vol(U)\vol(T)}\\
&\leq&\frac{w(U,T)}{\vol(U)}+\frac{8w(U,T)}{\varepsilon \vol(U)}\\
&\le&\frac{9}{\varepsilon}\frac{w(U,T)}{\vol(U)},
\end{eqnarray*}
which completes the proof.
\end{proof}

We are ready to prove the first part of   Theorem~\ref{thm:main}.

\begin{proof}[Proof of Theorem~\ref{thm:main}(i)]
For each $i\in[k]$, run Algorithm~\ref{alg:fair-trimming} on $A_i$.
The running time of the $i$th call is $\TildeO\left(\frac{|E(H_{A_i})|}{\eps}\right)$. We may omit zero-degree vertices from the auxiliary graphs, since they
contribute neither volume nor capacity. Since the sets $A_1,\ldots,A_k$
are pairwise disjoint, we have 
$
\sum_{i=1}^k |E(G[A_i])|\leq m$
and the total number of positive-degree vertices contained in the
$A_i$'s is at most $2m$. Hence, 
\[
\sum_{i=1}^k |E(H_{A_i})|\leq \sum_{i=1}^k |E(G[A_i])|+4m \leq  5m.
\]
Therefore, the total running time over all $k$ calls is
$\TildeO(m/\eps)$.

We now verify the claimed properties. By Lemma~\ref{lem:backbone-low}, every $B_i$ has positive volume and satisfies
\[
\phi_G(B_i)\leq \alpha p
=
\left(1+\frac{\eps}{16}\right)p.
\]
Thus, it remains to bound $\phi(G\{B_i\})$. We fix $i\in[k]$ and any $U\subseteq B_i$ satisfying  $0<\vol(U)\leq \frac{\vol(B_i)}{2}$. Let $T=B_i\setminus U$. By Corollary~\ref{cor:relative}, we have
\[\frac{w(U,T)}{\vol(U)}
\ge
\frac{\varepsilon}{9}
\max\{\phi_G(U),\phi_G(T)\}.\]
Notice that   the $k+1$ sets $A_1,\ldots,A_{i-1},U,T,A_{i+1},\ldots,A_k$ are pairwise disjoint, and have positive volume. Since
$
\phi_G(A_j)\leq p<\rho_G(k+1)$
for every $j\neq i$,
the definition of $\rho_G(k+1)$ implies
\[
\max\{\phi_G(U),\phi_G(T)\}\geq \rho_G(k+1).
\]
Hence
\[
\frac{w(U,B_i\setminus U)}{\vol(U)}
=
\frac{w(U,T)}{\vol(U)}
\geq
\frac{\eps\rho_G(k+1)}{9}.
\]
Since this holds for every $U\subseteq B_i$ with
$0<\vol(U)\leq\vol(B_i)/2$, we conclude that
\[
\phi(G\{B_i\})
\geq
\frac{\eps\rho_G(k+1)}{9}.
\]
This completes the proof.
\end{proof}

\subsection{Constructing a Partition}

In this subsection we prove that, given an input graph $G$ satisfying the condition of Theorem~\ref{thm:main} and the output of Algorithm~\ref{alg:fair-trimming}, one can construct a partition 
$V=P_1\sqcup\cdots\sqcup P_k$  in nearly-linear time.
We assume that $B_1,\ldots, B_k$ are the output   of Algorithm~\ref{alg:fair-trimming}. Hence, they are pairwise disjoint vertex sets with positive volumes satisfying $\phi_G(B_i)\leq\left(1+\eps/16\right)p$. To construct a partition $P_1\sqcup\cdots\sqcup P_k$, we build a balanced binary tree $\mathcal{T}$ inductively as follows:
\begin{enumerate}
    \item The root node $r$  is associated with the index set $I_r\leftarrow[k]$ and vertex set $X_r\leftarrow V$;
    \item For any node $u$ with $|I_u|\geq 2$, the algorithm splits $I_u$ into $I_{u_L}$ and $I_{u_R}$ such that $|I_{u_L} - I_{u_R}|\leq 1$. In the induced graph $G[X_u]$, the algorithm contracts
$\bigcup_{i\in I_{u_L}}B_i$ to $s$ and $\bigcup_{i\in I_{u_R}}B_i$ to $t$. Let $H$ be the resulting graph. The algorithm computes a $\left(1+1/\lceil\log_2k\rceil\right)$-fair $s,t$-cut in $H$,  and lets $X_{u_L}$ and $X_{u_R}$ be the two sides of the cut after undoing the contraction. These two sets $X_{u_L}\subsetneq V$ and $X_{u_R}\subsetneq V$ are the vertex sets associated with the two children of $u$.
\item The algorithm constructs the tree inductively as above, until a node $u$ corresponds to some index set $I_u$ of size $1$. 
The algorithm terminates on this branch, and sets $P_i\leftarrow X_u$.
\end{enumerate}
See Algorithm~\ref{alg:fair-attachment} for a formal description.  
Note that since all backbones indexed by $I_{u_L}$ are contracted into $s$ and all backbones indexed by $I_{u_R}$ are contracted into $t$, no backbone is ever split. Moreover, the two resulting child subsets $X_{u_L}$ and $X_{u_R}$ partition $X_u$, and  the sets $P_1,\ldots,P_k$ returned by Algorithm~\ref{alg:fair-attachment} satisfy
\[
V=P_1\sqcup\cdots\sqcup P_k,
\qquad
B_i\subseteq P_i
\quad\text{for every }i\in[k].
\]

\begin{algorithm}[ht]
\caption{\textsc{ConstructPartition}}
\label{alg:fair-attachment}
\begin{algorithmic}[1]
\Require Pairwise disjoint backbones $B_1,\ldots,B_k$
\Ensure A partition $V=P_1\sqcup\cdots\sqcup P_k$

\State Set
\[
h\gets\lceil\log_2 k\rceil,
\qquad
\alpha_a\gets1+\frac1h.
\]

\State Create a root node $r$ with
\[
I_r\gets[k],
\qquad
X_r\gets V.
\]

\While{there is a leaf node $u$ with $|I_u|\geq2$}
    \State Split $I_u=I_{u_L}\sqcup I_{u_R}$ so that
    \[
    \bigl||I_{u_L}|-|I_{u_R}|\bigr|\leq1.
    \]

    \State In $G[X_u]$, contract
    \[
    \bigcup_{i\in I_{u_L}}B_i
    \quad\text{to }s,
    \qquad
    \bigcup_{i\in I_{u_R}}B_i
    \quad\text{to }t.
    \]

    \State Compute an $\alpha_a$-fair $s,t$-cut $(S,T)$.

    \State Undo the contraction,  and let $X_{u_L}$ and $X_{u_R}$
    be the vertex sets corresponding to $S$ and $T$.

    \State Add $u_L$ and $u_R$ as the two children of $u$.
\EndWhile

\For{each leaf $u$ with $I_u=\{i\}$ for some $i\in [k]$}
    \[
    P_i\gets X_u.
    \]
\EndFor

\State \Return $P_1,\ldots,P_k$.
\end{algorithmic}
\end{algorithm}

The following lemma plays a key role in our analysis.

\begin{lemma}\label{lem:attachment}
Let $P_1\sqcup\cdots\sqcup P_k$ be the partition returned by Algorithm~\ref{alg:fair-attachment}. Then, it holds for any \(i\in [k]\) and   \(Y\subseteq P_i\setminus B_i\) that 
\[
w(Y,P_i\setminus Y)
\ge
\frac{1}{2\sqrt{\mathrm{e}}\,k}\cdot w(Y,V\setminus Y).
\]
\end{lemma}

\begin{proof}
Fix $i\in[k]$ and $Y\subseteq P_i\setminus B_i$. Let $u_0,u_1,\ldots,u_r$ be a path of $\mathcal{T}$ from the root node to a leaf  node labeled $i$, and for simplicity set $X_\ell\triangleq X_{u_\ell}$. Then,
\[
X_0=V\supsetneq X_1\supsetneq\cdots\supsetneq X_r=P_i,
\qquad
r\leq h.
\]
For $\ell=0,\ldots,r-1$, let
\[
Z_\ell\triangleq X_\ell\setminus X_{\ell+1},
\]
and $e_\ell\triangleq w(Y,Z_\ell)$. Finally, set $e_r\triangleq w(Y,P_i\setminus Y)$. 
Since $V=P_i\sqcup Z_0\sqcup\cdots\sqcup Z_{r-1}$, we have
\begin{equation}\label{eq:e-sum}
w(Y,V\setminus Y)
=
\sum_{\ell=0}^{r}e_\ell.
\end{equation}
Notice that, since $P_i$ contains $B_i$ and no other backbone and
$Y\subseteq P_i\setminus B_i$, the set $Y$ is disjoint from every
backbone. Hence, no vertex of $Y$ is contracted into a terminal vertex at any
level along a root-to-leaf path of $\mathcal{T}$.

Fix $\ell\in\{0,\ldots,r-1\}$, and consider the $\alpha_a$-fair cut
computed at node $u_\ell$. This cut separates the $X_{\ell+1}$ containing $P_i$ from $Z_\ell$. Suppose first that $X_{\ell+1}$ is the $t$-side of the cut. The cut edges from $Z_\ell$ into $Y$ have total capacity
\[
w(Y,Z_\ell)=e_\ell.
\]
By $\alpha_a$-fairness, these edges carry at least $e_\ell/\alpha_a$ units of flow from $Z_\ell$ into $Y$. Since $Y$ contains no terminal vertex, flow conservation implies that this net incoming flow leaves $Y$ through edges from $Y$ to $X_{\ell+1}\setminus Y$. By capacity feasibility, $e_\ell/\alpha_a\leq w(Y,X_{\ell+1}\setminus Y)$. If $X_{\ell+1}$ is instead the $s$-side of the cut, then fairness sends at least $e_\ell/\alpha_a$ units of flow from $Y$ toward $Z_\ell$. Flow conservation implies that the same amount of net flow   enters
$Y$ through edges between $Y$ and $X_{\ell+1}\setminus Y$. Thus the
same inequality holds. Therefore, in either case,
\begin{equation}\label{eq:attachment-level}
e_\ell
\leq
\alpha_a\,w(Y,X_{\ell+1}\setminus Y).
\end{equation}
As $X_{\ell+1}=P_i\sqcup Z_{\ell+1}\sqcup\cdots\sqcup Z_{r-1}$, we have
\[
w(Y,X_{\ell+1}\setminus Y)
=
e_r+\sum_{q=\ell+1}^{r-1}e_q
=
\sum_{q=\ell+1}^{r}e_q.
\]
Substituting this into \eqref{eq:attachment-level} gives
\begin{equation}\label{eq:attachment-rec}
e_\ell
\leq
\alpha_a\sum_{q=\ell+1}^{r}e_q.
\end{equation}
Let $F_\ell\triangleq \sum_{q=\ell}^{r}e_q$, for $\ell=0,\ldots,r$. By \eqref{eq:attachment-rec}, it holds for every $\ell<r$ that 
\[
F_\ell
=
e_\ell+F_{\ell+1}
\leq
(1+\alpha_a)F_{\ell+1},
\]
which implies 
\[
F_0
\leq
(1+\alpha_a)^rF_r
\leq
(1+\alpha_a)^h \cdot e_r.
\]
By \eqref{eq:e-sum}, we have
\[
w(Y,V\setminus Y)
\leq
(1+\alpha_a)^h\cdot  w(Y,P_i\setminus Y).
\]
Since $\alpha_a=1+\frac1h$, we have
\[
w(Y,V\setminus Y)\leq
\left(2+\frac1h\right)^h\cdot w(Y,P_i\setminus Y)
\leq
2^h\mathrm{e}^{1/2}\cdot w(Y,P_i\setminus Y)
\leq 2\mathrm{e}^{1/2}k\cdot w(Y,P_i\setminus Y),
\]
where the last inequality uses the fact that $h=\lceil\log_2k\rceil\leq \log_2k+1$. Therefore,
\[
w(Y,P_i\setminus Y)
\geq
\frac{1}{2\sqrt{\mathrm{e}}\,k}\cdot w(Y,V\setminus Y).
\]
This completes the proof.
\end{proof}

The following claim shows that every nonempty subset $S\subseteq P_i$ with $\vol(S\cap B_i)\leq\vol(B_i)/2$ has high conductance, and this will be used to lower bound the inner conductance of $G\{P_i\}$.

\begin{claim}\label{cl:key}
For every positive-volume $S\subsetneq P_i$ satisfying $\vol(S\cap B_i)\leq\vol(B_i)/2$, we have
\[
\frac{w(S,P_i\setminus S)}
{\min\{\vol(S),\vol(P_i\setminus S)\}}
\geq
\frac{\eps\rho_G(k+1)}{32\sqrt{\mathrm{e}}\,k}.
\]
\end{claim}

\begin{proof}
Let $S_B\triangleq S\cap B_i$ and $S_F \triangleq 
S\setminus B_i$. Suppose first that $S_B=\emptyset$. Since $P_i$ contains no backbone other than $B_i$, the $k+1$ sets $B_1,\ldots,B_k,S$ are pairwise disjoint. Moreover, Lemma~\ref{lem:backbone-low} gives $\phi_G(B_j)\leq(1+\eps/16)p<\rho_G(k+1)$ for every $j\in[k]$. Hence, by the definition of $\rho_G(k+1)$, we have $\phi_G(S)\geq\rho_G(k+1)$. Applying Lemma~\ref{lem:attachment} to $S\subseteq P_i\setminus B_i$ yields
\[
w(S,P_i\setminus S)
\geq
\frac{1}{2\sqrt{\mathrm{e}}\,k}w(S,V\setminus S)
\geq
\frac{\rho_G(k+1)}{2\sqrt{\mathrm{e}}\,k}\vol(S).
\]
Therefore,
\[
\frac{w(S,P_i\setminus S)}
{\min\{\vol(S),\vol(P_i\setminus S)\}}
\geq
\frac{\rho_G(k+1)}{2\sqrt{\mathrm{e}}\,k}
\geq
\frac{\eps\rho_G(k+1)}{32\sqrt{\mathrm{e}}\,k}.
\]
Hence, it suffices to analyze the case in which  $S_B$ has positive volume. Suppose first that $\vol(S_B)\geq\vol(S_F)$. The $k+1$ sets
$B_1,\ldots,B_{i-1},S_B,B_i\setminus S_B,B_{i+1},\ldots,B_k$
are pairwise disjoint and have positive volume. Since every $B_j$ for $j\neq i$ has conductance strictly smaller than $\rho_G(k+1)$, we have
$\max\{\phi_G(S_B),\phi_G(B_i\setminus S_B)\}\geq\rho_G(k+1)$. Hence, by  Corollary~\ref{cor:relative} we have 
\[
w(S_B,B_i\setminus S_B)
\geq
\frac{\eps\rho_G(k+1)}{9}\vol(S_B).
\]
Since all the edges between $S_B$ and $B_i\setminus S_B$  cross the cut $(S,P_i\setminus S)$ and $\vol(S)\leq2\vol(S_B)$, we have 
\[
\frac{w(S,P_i\setminus S)}
{\min\{\vol(S),\vol(P_i\setminus S)\}}
\geq
\frac{w(S,P_i\setminus S)}{\vol(S)}
\geq
\frac{\eps\rho_G(k+1)}{18},
\]
which is stronger than the claimed bound.

Finally, suppose that $\vol(S_F)\geq\vol(S_B)>0$. Since $\vol(S_B)\leq\vol(B_i)/2$, we have $\vol(B_i\setminus S_B)/\vol(B_i)\geq1/2$. Lemma~\ref{lem:one-shot} therefore gives
\[
w(S_B,B_i\setminus S_B)
\geq
\frac{\eps}{8}w(S_B,V\setminus B_i)
\geq
\frac{\eps}{8}w(S_B,S_F).
\]
  Consequently,
\begin{eqnarray*}
w(S,P_i\setminus S)
&\geq&
w(S_B,B_i\setminus S_B)
+w(S_F,P_i\setminus S_F)
-w(S_B,S_F)\\
&\geq&
w(S_F,P_i\setminus S_F)
-\left(1-\frac{\eps}{8}\right)w(S_B,S_F)\\
&\geq&
\frac{\eps}{8}w(S_F,P_i\setminus S_F),
\end{eqnarray*}
where the last inequality holds by the fact that $w(S_F,P_i\setminus S_F)\geq w(S_B,S_F)$. 
Since $S_F\subseteq P_i\setminus B_i$, the $k+1$ sets $B_1,\ldots,B_k,S_F$ are pairwise disjoint. As every $B_j$ has conductance strictly less than $\rho_G(k+1)$, we have  $\phi_G(S_F)\geq\rho_G(k+1)$. By Lemma~\ref{lem:attachment}, it holds that 
\[
w(S_F,P_i\setminus S_F)
\geq
\frac{\rho_G(k+1)}{2\sqrt{\mathrm{e}}\,k}\vol(S_F).
\]
Since $\vol(S)\leq2\vol(S_F)$, we obtain
\[
\frac{w(S,P_i\setminus S)}
{\min\{\vol(S),\vol(P_i\setminus S)\}}
\geq
\frac{w(S,P_i\setminus S)}{\vol(S)}
\geq
\frac{\eps\rho_G(k+1)}{32\sqrt{\mathrm{e}}\,k}.
\]
This proves the claim.
\end{proof}

Now we are ready to prove the second statement of Theorem~\ref{thm:main}.

\begin{proof}[Proof of Theorem~\ref{thm:main}(ii)]
We first analyze the time complexity of the algorithm. By the first statement of Theorem~\ref{thm:main}, we obtain backbones $B_1,\ldots, B_k$ in $\TildeO(m/\eps)$ time, and we analyze the time complexity of Algorithm~\ref{alg:fair-attachment}, which is dominated by computing the fair cuts. By \cite{LiLi24}, computing a $(1+\eta)$-fair cut takes $\TildeO(m/\eta)$ time and our algorithm computes an $\alpha_a$-fair cut for $\alpha_a=1+1/h$. Hence, it takes $\TildeO(h|E(G[X_u])|)$ time to compute a fair cut at node $u$. Since all the $X_u$'s in the same level of $\mathcal{T}$ are disjoint, we have $\sum_u|E(G[X_u])|\leq m$ and the algorithm's total running time at any fixed level is $\TildeO(mh)$. Since $\mathcal{T}$ has depth at most   $h=\lceil\log_2 k\rceil$, the total running time of Algorithm~\ref{alg:fair-attachment} is
$\TildeO(mh^2)=\TildeO(m\log^2 k)$.

Next we analyze the conductance of every $P_i$ in the partition.
Fix an arbitrary $i\in[k]$, and let $F_i\triangleq  P_i\setminus B_i$. Applying Lemma~\ref{lem:attachment} to $Y=F_i$ gives
\[
w(F_i,V\setminus P_i)
\leq
(2\sqrt{\mathrm{e}}\,k-1)\cdot w(F_i,B_i).
\]
Therefore,
\begin{eqnarray*}
w(P_i,V\setminus P_i)
&=&w(B_i,V\setminus P_i)+w(F_i,V\setminus P_i)\\
&\leq&w(B_i,V\setminus B_i)+(2\sqrt{\mathrm{e}}\,k-1)\cdot w(B_i,F_i)\\
&\leq&2\sqrt{\mathrm{e}}\,k\,w(B_i,V\setminus B_i).
\end{eqnarray*}
Since $\vol(P_i)\geq\vol(B_i)$, Lemma~\ref{lem:backbone-low} implies
\[
\phi_G(P_i)
\leq
2\sqrt{\mathrm{e}}\,k\,\phi_G(B_i)
\leq
2\sqrt{\mathrm{e}}\,k\left(1+\frac{\eps}{16}\right)p
\leq
\frac{17\sqrt{\mathrm{e}}}{8}kp
<
4kp.
\]

It remains to bound $\phi(G\{P_i\})$. 
For every nontrivial cut of $P_i$, at least one of its two sides $S$ satisfies $\vol(S\cap B_i)\leq\vol(B_i)/2$. We apply Claim~\ref{cl:key} to the side with volume at most $\vol(B_i)/2$, and obtain 
\[
\phi(G\{P_i\})
\geq
\frac{\eps\rho_G(k+1)}{32\sqrt{\mathrm{e}}\,k}
>
\frac{\eps\rho_G(k+1)}{53k}.
\]
This completes the proof.
\end{proof}

\section{Approximation Boost with Minimum-Cut Computation}\label{sec:epsapprox}

In this section we develop a general algorithmic framework called \textsc{ApproximationBoost} that improves the outer conductance of any $k$ disjoint subsets of positive volume under an assumption between $\rho_G(k+1)$ and $\rho_G(k)$, and prove Theorem~\ref{thm:epsapprox}. Combining this with Theorem~\ref{thm:main}, we obtain Corollary \ref{cor:largegap}.

\subsection{Algorithm}

For any $k$ pairwise disjoint and nonempty sets $B_1,\ldots, B_k$ and parameter $\alpha\geq 1$ as input, our designed algorithm~\textsc{ApproximationBoost} consists of $O(\log\alpha)$ iterations. In the $r$-th iteration, the algorithm performs the following:
\begin{enumerate}
    \item The algorithm sets $h\triangleq\max_{i\in[k]} \phi_G(B_i)$, and  \[
    q_r\triangleq \frac{2h}{\alpha}\cdot \left(\frac{3}{2}\right)^r;
    \]
    \item For every iteration $r$  and every $i\in [k]$, the algorithm performs the following:
     \begin{enumerate}
     \item The algorithm constructs graph $H^{(r)}_{i}$ as follows: 
     \begin{enumerate}
           \item The vertex set of $H^{(r)}_i$ consists of all the vertices in $G$ and two additional ones $s,t$;
           \item Every edge $\{u,v\}$ in  $G$ is included in $H^{(r)}_{i}$ with capacity $w(u,v)$;
          \item For every vertex $u\in B_i$, edge $\{s, u\}$ is included in $H^{(r)}_{i}$ with capacity $q_r\deg(u)$;
          \item For every vertex $u\not\in B_i$, edge $\{u,t\}$ is included in $H^{(r)}_{i}$ with capacity $q_r\deg(u)$;
    \end{enumerate}
    \item The algorithm computes a minimum  $s$-$t$ cut $\left(S^{(r)}_{i}, T^{(r)}_{i}\right)$ in $H^{(r)}_{i}$, and sets $D^{(r)}_{i}\triangleq S^{(r)}_{i}\setminus \{s\}$;
    \item The algorithm computes $C_i^{(r)}\gets D_i^{(r)}
        \setminus\left(\bigcup_{j\in[k]\setminus\{i\}}D_j^{(r)}\right)$,
        and stores the current $C^{(r)}_1,\ldots, C^{(r)}_k$ if $\max_{i\in[k]}\phi_G\left(C^{(r)}_i\right)$ is smaller than the previously recorded one. 
\end{enumerate}
\item The algorithm returned the stored subsets $C^{(j)}_1,\ldots, C^{(j)}_k$ for some $j$.
\end{enumerate}
See Algorithm~\ref{alg:eps-approx} for a formal description.
\begin{algorithm}[!ht]
\caption{\textsc{ApproximationBoost}
}
\label{alg:eps-approx}
\begin{algorithmic}[1]
\Require an undirected graph $G=(V,E,w)$, an integer $k$,
pairwise disjoint and nonempty sets $B_1,\ldots,B_k$,
and parameter $\alpha\geq 1$
\Ensure Pairwise disjoint and  nonempty sets $C_1,\ldots,C_k$

\State $h\gets\max_{i\in[k]}\phi_G(B_i)$
\State $\mathrm{best}\gets+\infty$
\State $\mathcal{C}^*\gets\mathrm{null}$

\For{$r=0,1,\ldots,\lceil\log_{\frac{3}{2}}\alpha\rceil$}
    \State $q_r\gets \frac{2h}{\alpha}\cdot \left(\frac{3}{2}\right)^r$

    \For{$i=1,2,\ldots,k$}
        \State $D_i^{(r)}\gets
        \Call{PenalizedMinCut}{G,B_i,q_r}$
    \EndFor

    \For{$i=1,2,\ldots,k$}
        \State $C_i^{(r)}\gets D_i^{(r)}
        \setminus\left(\bigcup_{j\in[k]\setminus\{i\}}D_j^{(r)}\right)$
    \EndFor

    \If{$\vol(C_i^{(r)})>0$ for all $i\in[k]$}
        \State $\gamma_r\gets
        \max_{i\in[k]}\phi_G(C_i^{(r)})$
        \If{$\gamma_r<\mathrm{best}$}
            \State $\mathrm{best}\gets\gamma_r$
            \State $\mathcal{C}^*\gets
            (C_1^{(r)},\ldots,C_k^{(r)})$
        \EndIf
    \EndIf
\EndFor

\State \Return $\mathcal{C}^*$

\vspace{1mm}
\Function{PenalizedMinCut}{$G,B,q$}
    \State Construct an undirected auxiliary graph $H$
    with vertices $V\cup\{s,t\}$
    \ForAll{$\{u,v\}\in E$}
        \State Add edge $\{u,v\}$ with capacity $w(u,v)$
    \EndFor
    \ForAll{$v\in V$}
        \If{$v\in B$}
            \State Add edge $\{s,v\}$ with capacity $q\deg(v)$
        \Else
            \State Add edge $\{v,t\}$ with capacity $q\deg(v)$
        \EndIf
    \EndFor
    \State Compute a minimum $s$-$t$ cut $(S,T)$ in $H$ with $s\in S$ and $t\in T$
    \State \Return $S\setminus\{s\}$
\EndFunction
\end{algorithmic}
\end{algorithm}

\subsection{Analysis}

Now we analyze Algorithm~\ref{alg:eps-approx}, and prove Theorem~\ref{thm:epsapprox}. We first introduce the notion of central disjoint subsets.

\begin{definition}[central disjoint subsets]
For any weighted graph $G=(V,E,w)$ and some positive integer $k$, we call $k$ pairwise disjoint subsets $A_1, \dots, A_k$ \emph{central} if (i)   $\max_{i\in[k]} \phi_G(A_i)=\rho_G(k)$, and (ii) for every $i\in [k]$ there is no proper subset $S\subsetneq A_i$ with $\phi_G(S)\leq \phi_G(A_i)$. 
\end{definition}

The following lemma shows that the inner conductance of any central disjoint subsets can be lower bounded with respect to the difference between $\rho_G(k+1)$ and $\rho_G(k)$. Moreover, for any subset $B\subseteq V$ that doesn't have large overlap with any central disjoint subsets, $\phi_G(B)$ can be lower bounded with respect to the difference between $\rho_G(k+1)$ and $\rho_G(k)$ as well.

\begin{lemma}\label{lem:centralitygivesgoodexpanders}
Let $G=(V,E,w)$ and $A_1,\dots,A_k$ be central
$k$ disjoint subsets. If $\rho_G(k+1)>\rho_G(k)$, then the following holds:
\begin{itemize}
    \item [\textup{(i)}] It holds for any $i\in[k]$ that 
    \[\phi(G\{A_i\})>
    \frac{\rho_G(k+1)-\rho_G(k)}{2}; \]
    \item [\textup{(ii)}] For any $B\subseteq V$ with
    $\vol(B)>0$ and
    $\vol(B\cap A_j)\leq\vol(A_j)/2$
    for all $j\in[k]$, it holds  that
    $\phi_G(B)\geq\rho_G(k+1)-2\rho_G(k)$.
\end{itemize}
\end{lemma}

\begin{proof}
By centrality,  it holds for any positive volume subset $X\subsetneq A_i$ that 
$\phi_G(A_i)<\phi_G(A_i\setminus X)$, which implies
\[
w(A_i,V\setminus A_i)(\vol(A_i)-\vol(X))
<
(w(A_i\setminus X,V\setminus A_i)
+w(A_i\setminus X,X))\vol(A_i),
\]
and therefore,
\begin{equation}\label{eq:central-boundary}
w(X,V\setminus A_i)\leq w(X,A_i\setminus X)+\rho_G(k)\vol(X).
\end{equation}
Note that the inequality in \eqref{eq:central-boundary}
is strict for every nonempty proper subset $X\subsetneq A_i$.
It is not hard to check that \eqref{eq:central-boundary}
also holds when $X=\emptyset$ or $X=A_i$.

To prove (i), we choose any nonempty proper subset
$X\subsetneq A_i$ with
$\vol(X)\leq\vol(A_i)/2$.
Since $A_1,\dots,A_{i-1},X,A_i\setminus X,
A_{i+1},\dots,A_k$ are $k+1$ disjoint
subsets, we have that
$\max\{\phi_G(X),\phi_G(A_i\setminus X)\}
\geq \rho_G(k+1)$.
Applying \eqref{eq:central-boundary} to the
subset with conductance at least $\rho_G(k+1)$, we obtain
\[
w(X,A_i\setminus X)>
\frac{\rho_G(k+1)-\rho_G(k)}{2}
\min\{\vol(X),\vol(A_i\setminus X)\}
=\frac{\rho_G(k+1)-\rho_G(k)}{2}\vol(X),
\]
which proves (i).

Next, we prove (ii). Without loss of generality, we assume $\rho_G(k+1)>2\rho_G(k)$, since otherwise the statement is trivial. 
Suppose for a contradiction that there is some $B\subseteq V$ with $\vol(B)>0$, such that $\vol(B\cap A_j)\leq \vol(A_j)/2$ for all $j\in[k]$ and $\phi_G(B)<\rho_G(k+1)-2\rho_G(k)$.
Let $J\triangleq \{j\in [k]:\phi_G(A_j\setminus B)\geq\rho_G(k+1)\}$, and  $B' \triangleq B\setminus \left(\bigcup_{j\in J}A_j\right)$. We now claim that
$\vol(B')>0$ and
\[\phi_G(B') < \rho_G(k+1)-2\rho_G(k).\]
We prove this by induction on the number of $A_j$s
that have been removed from $B$.
Let $J'\subseteq J$ and
$T= B\setminus \left(\bigcup_{j\in J'}A_j\right)$.
We prove the stronger claim that
$w(T,V\setminus T)<(\rho_G(k+1)-2\rho_G(k))\vol(T)$.
The conclusion holds when $|J'|=0$, since 
$\phi_G(B)<\rho_G(k+1)-2\rho_G(k)$
and $\vol(B)>0$.

Suppose the conclusion holds when
$|J'|=t\geq 0$, and we prove
  that it holds for $|J'|=t+1$. 
Assume that $T=T'\setminus A_{j^*}$ with
$j^*\in J$ and $T'$ is obtained from $B$ by
removing $t$ $A_j$s from $J\setminus \{j^*\}$.
Then, by the induction hypothesis, we have that
\[
w(T',V\setminus T')<(\rho_G(k+1)-2\rho_G(k))\vol(T').
\]
By the posimodularity of the cut function,
we have
\begin{eqnarray*}
\lefteqn{w(A_{j^*}\setminus B, V\setminus (A_{j^*}\setminus B))+w(T,V\setminus T)}\\
&=&w(A_{j^*}\setminus T',V\setminus(A_{j^*}\setminus T'))+w(T'\setminus A_{j^*},V\setminus (T'\setminus A_{j^*}))\\
    &\leq& w(A_{j^*},V\setminus A_{j^*})+w(T',V\setminus T').
\end{eqnarray*}
Thus, by the induction hypothesis and the facts that
$j^*\in J$, $\phi_G(A_{j^*})\leq \rho_G(k)$
and $2\vol(A_{j^*}\setminus B)\geq \vol(A_{j^*})$,
we have that
\begin{eqnarray*}
\rho_G(k+1)\vol(A_{j^*}\setminus B)+w(T,V\setminus T)&<& \rho_G(k)\vol(A_{j^*})+(\rho_G(k+1)-2\rho_G(k))\vol(T')\\
    &\leq& 2\rho_G(k)\vol(A_{j^*}\setminus B)+(\rho_G(k+1)-2\rho_G(k))\vol(T').
\end{eqnarray*}
Combining this with  $\vol(A_{j^*}\cap B)\leq \vol(A_{j^*}\setminus B)$ implies that 
\begin{eqnarray*}
    \lefteqn{w(T,V\setminus T)}\\
    &<& -(\rho_G(k+1) - 2\rho_G(k))\vol(A_{j^*}\setminus B)+(\rho_G(k+1)-2\rho_G(k))(\vol(T)+\vol(A_{j^*}\cap B))\\
    &\leq & (\rho_G(k+1)-2\rho_G(k))\vol(T),
\end{eqnarray*}
which proves the claim.  

Now consider the $k+1$ disjoint subsets
$B'$, $A_j$ with $j\in J$,
and $A_j\setminus B$ for the remaining
$j\in [k]\setminus J$.
These $k+1$ subsets are nonempty, have positive volume, and have   conductance strictly less than $\rho_G(k+1)$, which leads to a  contradiction. This proves (ii).
\end{proof}

The next lemma plays a key role in our analysis and proves that, if $\rho_G(k+1)/\rho_G(k)$ is large, then every family of $k$ disjoint subsets $B_1,\ldots,B_k$ with a low value of $\rho(k)$ has large overlap with central disjoint subsets $A_1,\ldots, A_k$; i.e., there is a permutation $\pi:[k]\mapsto[k]$ such that $\vol(B_i\triangle A_{\pi(i)})$ is low. We will present a family of graphs in Section~\ref{sec:tightness} to demonstrate that the   bound below is tight up to constant factors.

\begin{lemma}\label{lem:volconcentration}
    Let $G=(V,E,w)$ be an undirected graph satisfying $\rho_G(k+1)/\rho_G(k)>\alpha+2$ for some $\alpha\geq 1$ and $\rho_G(k)>0$. Let $A_1,\dots,A_k$ be central $k$ pairwise disjoint subsets of $G$, and $B_1,\dots,B_k$ be $k$ pairwise disjoint subsets with $\max_{i\in[k]}\phi_G(B_i)\leq \alpha\rho_G(k)$. Then, there exists a permutation $\pi:[k]\mapsto[k]$ such that for every $i\in [k]$,
    \[
    \vol(B_i\triangle A_{\pi(i)})\leq
    \frac{3\alpha+1}{\rho_G(k+1)/\rho_G(k)-\alpha-2}
    \cdot\vol(A_{\pi(i)}).
    \]
\end{lemma}

\begin{proof}
    Throughout this proof we set  $g=\rho_G(k+1)/\rho_G(k)$.
    We first show that, for every $i\in[k]$, there exists a $j\in[k]$ such that $\vol(B_i\cap A_j)>\vol(A_j)/2$. Assume for contradiction that there is some $i$ such that $\vol(B_i\cap A_j)\leq\vol(A_j)/2$ for all $j\in[k]$. Then, by the second statement of  Lemma \ref{lem:centralitygivesgoodexpanders}, we have that
    \[\phi_G(B_i)\geq \rho_G(k+1)-2\rho_G(k)=(g-2)\rho_G(k)>\alpha\rho_G(k),\] 
    where the last inequality holds as $g>\alpha+2$, contradicting $\max_{i\in[k]}\phi_G(B_i)\leq \alpha\rho_G(k)$.

Hence, for every $i\in[k]$ there exists a $j\in[k]$ such that $\vol(B_i\cap A_j)>\vol(A_j)/2$.
As the $B_i$s are disjoint subsets, each $A_j$ can have such an overlap with at most one $B_i$. Since there are $k$ subsets on each side,  for every $i\in[k]$ there is a unique $j$ such that $\vol(B_i\cap A_j)>\vol(A_j)/2$. We define a function $\pi:[k]\mapsto [k]$ by $\pi(i)=j$, and it's easy to see that   $\pi:[k]\mapsto [k]$ is a permutation.

Now we prove that $\pi$ satisfies the claimed property. Fix an $i\in [k]$. By definition, we have that $\vol(B_i\cap A_{\pi(i)})>\vol(A_{\pi(i)})/2$ and $\vol(B_i\cap A_j)\leq\vol(A_j)/2$ for all $j\in[k]\setminus\{\pi(i)\}$. Let $X=B_i\setminus A_{\pi(i)}$ and $Y=B_i\cap A_{\pi(i)}$. We therefore have for every $j\neq \pi(i)$ that 

\begin{equation}\label{eq:s5l21}
    \vol(X\cap A_j)\leq \vol(A_j)-\vol(A_j\cap B_{\pi^{-1}(j)})\leq \vol(A_j)/2,
\end{equation}
and
\begin{equation}\label{eq:s5l22}
    \vol(X\cap A_{\pi(i)})=\vol(\emptyset)=0\leq \vol(A_{\pi(i)})/2.
\end{equation}
In addition, by \eqref{eq:central-boundary}, we have that
\[
w(X,Y)\leq w(Y,V\setminus A_{\pi(i)})
\leq w(Y,A_{\pi(i)}\setminus B_i)
+\rho_G(k)\vol(Y).
\]
Thus,
\begin{eqnarray*}
w(X,V\setminus X) &=& w(X,V\setminus B_i)+w(X,Y)\\ 
&\leq& w(X,V\setminus B_i)+w(Y,A_{\pi(i)}\setminus B_i)+\rho_G(k)\vol(Y)\\
&\leq& w(B_i,V\setminus B_i)+\rho_G(k)\vol(Y)\\
&\leq& \alpha\rho_G(k)\vol(X)+(\alpha+1)\rho_G(k)\vol(Y),
\end{eqnarray*}
where the last inequality holds as $\phi_G(B_i)\leq \alpha\rho_G(k)$. Thus, by \eqref{eq:s5l21},   \eqref{eq:s5l22}, and the second statement of Lemma \ref{lem:centralitygivesgoodexpanders},  we have  $\phi_G(X)\geq (g-2)\rho_G(k)$, which implies

\[\left(g-2\right)\rho_G(k)\vol(X)\leq w(X,V\setminus X)\leq \alpha\rho_G(k)\vol(X)+(\alpha+1)\rho_G(k)\vol(Y),\]
and therefore,
\begin{equation}\label{eq:s5l11}
    \vol(B_i\setminus A_{\pi(i)})
    \leq\frac{\alpha+1}{g-\alpha-2}
    \vol(B_i\cap A_{\pi(i)}).
\end{equation}
On the other hand, by the first statement of  Lemma \ref{lem:centralitygivesgoodexpanders}, as $\vol(A_{\pi(i)}\setminus B_i)\leq \vol(A_{\pi(i)})/2$, we have that
\begin{equation}\label{eq:s5l12}
    \vol(A_{\pi(i)}\setminus B_i)
    \leq\frac{2}{\rho_G(k+1)-\rho_G(k)}
    w(B_i\cap A_{\pi(i)},A_{\pi(i)}\setminus B_i)
    \leq\frac{2\alpha}{g-1}\vol(B_i).
\end{equation}
By \eqref{eq:s5l11}, we also have
$\vol(B_i)\leq\frac{g-1}{g-\alpha-2}
\vol(B_i\cap A_{\pi(i)})$.
Thus, combining \eqref{eq:s5l11} and \eqref{eq:s5l12}, we obtain
\[
\begin{aligned}
\vol(B_i\triangle A_{\pi(i)})
&=\vol(B_i\setminus A_{\pi(i)})
+\vol(A_{\pi(i)}\setminus B_i)\\
&\leq\left(
\frac{\alpha+1}{g-\alpha-2}
+\frac{2\alpha}{g-1}\cdot\frac{g-1}{(g-\alpha-2)}
\right)\vol(B_i\cap A_{\pi(i)})\\
&\leq\frac{3\alpha+1}{g-\alpha-2}\cdot
\vol(A_{\pi(i)}),
\end{aligned}
\]
which completes the proof.
\end{proof}

Now we are ready to prove Theorem \ref{thm:epsapprox}.

\begin{proof}[Proof of Theorem \ref{thm:epsapprox}.] The number of minimum $s$-$t$ cut computations claimed in the theorem follows directly from Algorithm~\ref{alg:eps-approx}, so we only need to prove the approximation guarantee of the algorithm.  Suppose that $C_1^{(r^*)},\ldots,C_k^{(r^*)}$
are the output of  Algorithm \ref{alg:eps-approx}.
Since $\max_{i\in[k]}\phi_G(B_i)=h
\leq\alpha\rho_G(k)$, it holds that 
$2h/\alpha\leq2\rho_G(k)\leq2h$. By the definition of 
\[
q_r\triangleq \frac{2h}{\alpha}\cdot \left(\frac{3}{2}\right)^r,
\]
we have     
\[
q_0= \frac{2h}{\alpha}\leq2\rho_G(k)
\qquad\text{and}\qquad
q_{\lceil\log_{\frac32}\alpha\rceil}
\geq2h\geq2\rho_G(k),
\]
which, combined with the fact that
$q_{r+1}/q_r=3/2$, implies that there is some
$r$ with $\tau\triangleq q_r/\rho_G(k)\in[2,3)$.
Fix such $r$, and let $C_1,\dots,C_k$ be
the $k$ subsets obtained in the iteration
corresponding to $q_r$. We will show that these subsets are nonempty and satisfy the desired approximation guarantee.

For any fixed $i\in[k]$, it is not hard to show,
from how we design
$\Call{PenalizedMinCut}{G,B_i,q_r}$,
that $D_i^{(r)}$ minimizes the following
function among all $X_i\subseteq V$
\[
f_i(X_i)=w(X_i,V\setminus X_i)
+q_r\vol(B_i\triangle X_i).
\]
We claim that $C_i$ also minimizes $f_i$.
Indeed, for any $i\neq j$ and minimizers $X$
and $Y$ of $f_i$ and $f_j$, respectively,
the posimodularity of the cut function and
$B_i\cap B_j=\emptyset$ imply that
$f_i(X\setminus Y)+f_j(Y\setminus X)
\leq f_i(X)+f_j(Y)$.
By optimality, $X\setminus Y$ is also a
minimizer of $f_i$. Applying this argument
successively to $D_i^{(r)}$ and $D_j^{(r)}$
for all $j\neq i$ shows that $C_i$ minimizes
$f_i$. Moreover, the $C_i$s are disjoint
by construction.

Let $A_1,\dots,A_k$ be central $k$ disjoint
subsets of $G$, and
\[
\delta\triangleq\frac{3\alpha+1}
{\rho_G(k+1)/\rho_G(k)-\alpha-2}.
\]
Then, by Lemma \ref{lem:volconcentration} 
we can assume without loss of generality that
\begin{equation}\label{eq:s5t10}
    \vol(B_i\triangle A_i)
    \leq\delta\cdot\vol(A_i).
\end{equation}
Since
$\max_{i\in[k]}\phi_G(A_i)=\rho_G(k)$,
we have that
\begin{equation}\label{eq:s5t11}
    w(C_i,V\setminus C_i)
    +q_r\vol(B_i\triangle C_i)
    =f_i(C_i)\leq f_i(A_i)
    \leq\rho_G(k)\vol(A_i)
    +q_r\vol(B_i\triangle A_i)
\end{equation}
and in particular
\begin{equation}\label{eq:s5t12}
    \vol(B_i\triangle C_i)
    \leq(\rho_G(k)/q_r+\delta)\vol(A_i)
    =(1/\tau+\delta)\vol(A_i).
\end{equation}
By the triangle inequality of symmetric differences,
we have that
\begin{equation}\label{eq:s5t13}
    \vol(C_i)
    \geq\vol(A_i)-\vol(C_i\triangle A_i)
    \geq\vol(A_i)-\vol(B_i\triangle A_i)
    -\vol(B_i\triangle C_i)
\end{equation}
Thus, on the one hand, by \eqref{eq:s5t10}
and \eqref{eq:s5t12} and the fact that
$\tau\geq2$ we have that
\begin{equation}\label{eq:s5t14}
    \vol(C_i)
    \geq(1-1/\tau-2\delta)\vol(A_i),
\end{equation}
which is positive as $\tau\geq2$ and
$\delta<1/4$. We remark that $\delta<1/4$  holds since
\[
\rho_G(k+1)/\rho_G(k)\geq 
\frac{18(\alpha+1)(1+\eps)}{\eps}>18(\alpha+1)
\]
by the condition of Theorem~\ref{thm:epsapprox}.

On the other hand, we have that
\begin{eqnarray*}
    &~&w(C_i,V\setminus C_i)
    +\rho_G(k)(\vol(A_i)-\vol(C_i))\\
    &\leq&
    w(C_i,V\setminus C_i)
    +(q_r-\rho_G(k))\vol(B_i\triangle C_i)
    +\rho_G(k)(\vol(A_i)-\vol(C_i))\\
    &\leq &
    w(C_i,V\setminus C_i)
    +(q_r-\rho_G(k))\vol(B_i\triangle C_i)
    +\rho_G(k)(\vol(B_i\triangle C_i)
    +\vol(B_i\triangle A_i))\\
    &=&
    w(C_i,V\setminus C_i)
    +q_r\vol(B_i\triangle C_i)
    +\rho_G(k)\vol(B_i\triangle A_i)\\
    &\leq &
    \rho_G(k)\vol(A_i)
    +(q_r+\rho_G(k))\vol(B_i\triangle A_i)\\
    &\leq &
    (\rho_G(k)+(q_r+\rho_G(k))\delta)\vol(A_i)\\
    &\leq &
    \rho_G(k)\vol(A_i)
    +\frac{(\tau+1)\delta\rho_G(k)}
    {1-1/\tau-2\delta}\vol(C_i),
\end{eqnarray*}
where the second inequality follows by \eqref{eq:s5t13}, the third one follows by  \eqref{eq:s5t11}, the fourth one follows by \eqref{eq:s5t10}, and the final one follows by 
\eqref{eq:s5t14}. Moreover, since it holds for $\tau\in [2,3)$ that
\[
3\left(1-\frac1\tau-2\delta\right)
-(\tau+1)\left(\frac12-2\delta\right)
=\frac{(\tau-2)(3-\tau)}{2\tau}
+2(\tau-2)\delta
\geq0, 
\]
we have that
\[
\phi_G(C_i)
=\frac{w(C_i,V\setminus C_i)}{\vol(C_i)}
\leq\left(1+
\frac{(\tau+1)\delta}{1-1/\tau-2\delta}\right)\rho_G(k)
\leq\left(1+
\frac{3\delta}{1/2-2\delta}\right)\rho_G(k)
\leq(1+\eps)\rho_G(k);
\]
here, the last inequality holds since
$\delta\leq\frac{\eps}{2(3+2\eps)}$,
which can be derived by the condition that
\[
\rho_G(k+1)/\rho_G(k)\geq 
\frac{18(\alpha+1)(1+\eps)}{\eps}.
\]
Hence, the candidate corresponding to $q_r$
consists of $k$ disjoint nonempty subsets
satisfying the desired guarantee.
As Algorithm \ref{alg:eps-approx} returns
the valid candidate with the lowest maximum
conductance, we have that
\[
\max_{i\in[k]}\phi_G\left(C_i^{(r^*)}\right)
\leq\max_{i\in[k]}\phi_G(C_i)
\leq(1+\eps)\rho_G(k),
\]
which completes the proof.
\end{proof}

Now we give the proof for Corollary \ref{cor:largegap}.

\begin{proof}[Proof of Corollary~\ref{cor:largegap}.]
For any $0<\eps\leq 1$, by Theorem \ref{thm:epsapprox},
in $O\left(k(1+\log\alpha)m^{1+o(1)}\right)$ time
we obtain $k$ disjoint subsets $A_1,\dots,A_k$ with
$\max_{1\leq i\leq k}\phi_G(A_i)\leq
(1+\eps)\rho_G(k)$.
Since $\alpha\geq 1$ and $\eps\leq 1$, we have
\[
\rho_G(k+1)
\geq \frac{18(\alpha+1)(1+\eps)}{\eps}\cdot\rho_G(k)
\geq 36(1+\eps)\rho_G(k)
\geq 2(1+\eps)\rho_G(k)=(1+1)(1+\eps)\rho_G(k).
\]
Thus, applying Theorem \ref{thm:main} to these subsets
with $p=(1+\eps)\rho_G(k)$ and its gap parameter
set to $1$ gives the results. Since the gap parameter in Theorem \ref{thm:main} is set to $1$, the additional running time is $\widetilde O(m\log^2 k)$, which is absorbed
into the stated time bound.
\end{proof}

\subsection{On the Tightness of Lemma~\ref{lem:volconcentration}\label{sec:tightness}}

Finally, at the end of the section we present a class of graphs showing that  our presented bound in Lemma~\ref{lem:volconcentration} is tight up to a constant. 

\begin{proposition}\label{prop:concentrationtight}
The bound in Lemma \ref{lem:volconcentration}
is asymptotically tight up to a constant factor.
More precisely, for every two integers $q\geq20$
and $1\leq a\leq q$, there exists an unweighted
simple connected graph $G$ with
$g\triangleq \rho_G(3)/\rho_G(2)=\Theta(q^2)$,
central $2$ disjoint subsets $A_1,A_2$,
and two disjoint subsets $B_1,B_2$ satisfying
$\max_{i\in[2]}\phi_G(B_i)\leq a\rho_G(2)$,
such that for every permutation $\pi:[2]\mapsto[2]$,
\[
\max_{i\in[2]}
\frac{\vol(B_i\triangle A_{\pi(i)})}
{\vol(A_{\pi(i)})}
=\Omega\left(\frac{a}{g}\right).
\]
\end{proposition}

\begin{figure}[ht]
\centering
\begin{tikzpicture}[
    scale=0.9,
    every node/.style={font=\small},
    clique/.style={draw=black, thick, fill=gray!12},
    vertex/.style={
        circle, draw=black, fill=white,
        inner sep=1.6pt
    }
]

\draw[clique] (0,0) ellipse (1.35 and 0.95);
\node at (-0.2,0.15) {$U=K_{2q}$};

\draw[clique] (6,0) ellipse (1.05 and 0.85);
\node at (6.2,0.15) {$W=K_q$};

\node[vertex] (u) at (1.3,0) {};
\node[vertex] (v) at (5,0) {};
\node[left] at (1.3,-0.1) {$u$};
\node[right] at (5,-0.1) {$v$};
\draw[thick] (u) -- (v);

\node[vertex] (z1) at (1.9,1.45) {};
\node[vertex] (z2) at (2.7,1.65) {};
\node[vertex] (za) at (3.8,1.45) {};

\draw[thick] (u) -- (z1);
\draw[thick] (u) -- (z2);
\draw[thick] (u) -- (za);

\node[above] at (z1) {$z_1$};
\node[above] at (z2) {$z_2$};
\node[above] at (za) {$z_a$};
\node at (3.3,1.7) {$\cdots$};

\node at (0,-1.5)
    {$A_1=U\cup Z,\quad B_1=U$};
\node at (6,-1.5)
    {$A_2=B_2=W$};

\end{tikzpicture}
\caption{An unweighted graph showing the tightness
of Lemma~\ref{lem:volconcentration}.
Here, $Z=\{z_1,\ldots,z_a\}$ consists of
$a$ pendant vertices adjacent to the same
vertex $u\in U$. Edges within the two
cliques are omitted.}
\label{fig:concentration-tight}
\end{figure}
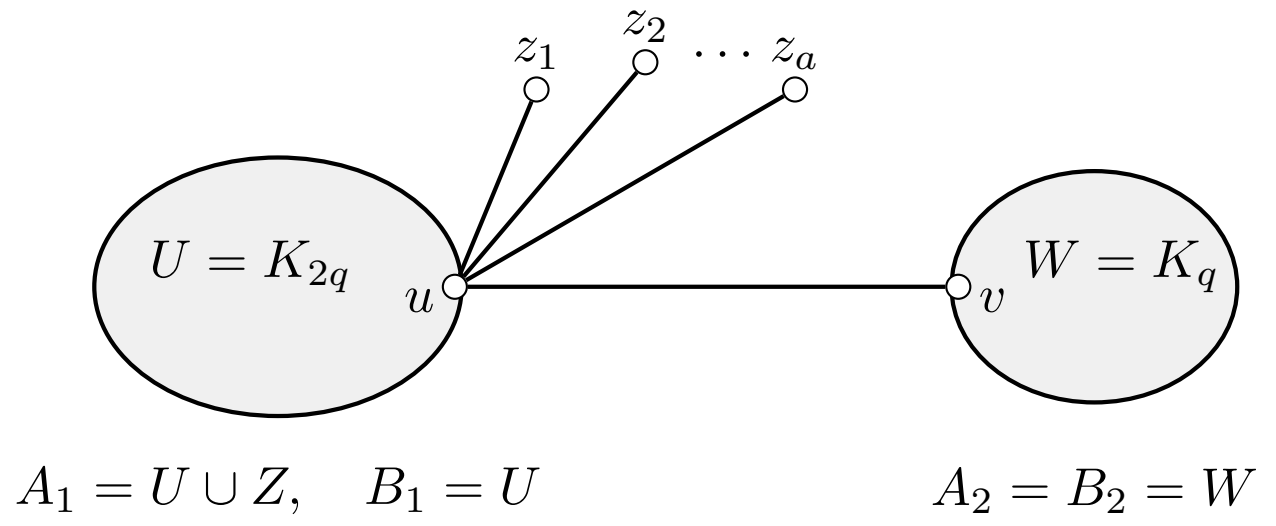

\begin{proof}
For any integers $q\geq20$ and $1\leq a\leq q$,
let $G$ be the graph constructed as follows: (1) $G$ consists of $3q+a$ vertices; (2) the first $2q$ vertices of $G$ form a complete graph $U=K_{2q}$, and another $q$ vertices form the other complete graph $W=K_q$. Moreover, we pick a vertex $u$ from $K_{2q}$,   a vertex $v$ from $K_q$, and add $\{u,v\}$ to the edge set of $G$; (3) we further connect $u$ to $a$ additional vertices $z_1,\ldots,z_a$. See Figure~\ref{fig:concentration-tight} for illustration.

We first prove that
\[
\rho_G(2)=\frac{1}{q(q-1)+1}
\quad\text{and}\quad
\frac14\leq\rho_G(3)\leq1.
\]
Notice that any cut  splitting $U$ or $W$
has at least $q-1$ edges along the cut, and hence
has conductance greater than
$1/(q(q-1)+1)$, since
\[
\frac{q-1}{\vol(V)}>
\frac{1}{q(q-1)+1}
\]
for $q\geq20$ and $a\leq q$.
Thus, if there were two disjoint subsets
with conductance strictly less than
$1/(q(q-1)+1)$, they would have to contain
$U$ and $W$ entirely, respectively, since
a subset containing only pendant vertices
has conductance $1$.
However, any subset containing $W$ but
not $U$ has conductance at least
$1/(q(q-1)+1)$.
Since
$\phi_G(W)=1/(q(q-1)+1)$ and
$\phi_G(U\cup Z)<\phi_G(W)$,
we have that
$\rho_G(2)=1/(q(q-1)+1)$.

Next, we analyze the range of $\rho_G(3)$. We note that any subset $S$
containing at most half of the vertices
of both $U$ and $W$ has conductance
at least $1/4$.
Indeed, let $x,y,z$ be the numbers of
vertices of $S$ in $U,W,Z$,
respectively. If the vertex $u$ of $U$
adjacent to $Z$ belongs to the subset,
then
\[
w(S,V\setminus S)\geq qx+\frac{q}{2}y,
\qquad
\vol(S)\leq4qx+qy.
\]
Otherwise,
\[
w(S,V\setminus S)\geq qx+\frac{q}{2}y+z,
\qquad
\vol(S)\leq2qx+qy+z.
\]
Thus, $\phi_G(S)\geq1/4$ in both cases.
As there cannot be three disjoint subsets
each containing more than half of $U$ or $W$,
we have $\rho_G(3)\geq1/4$.
The upper bound follows by taking three
singleton subsets. Consequently,
$g=\Theta(q^2)$.

Now let $A_1=U\cup Z$, $A_2=W$,
$B_1=U$ and $B_2=W$.
Since
\[
\phi_G(A_1)
=\frac{1}{2q(2q-1)+1+2a}
<\phi_G(A_2)=\rho_G(2),
\]
the subsets $A_1,A_2$ are the optimal sets corresponding to $\rho_G(2)$.
Moreover, they are central.
Indeed, any nonempty proper subset of $A_1$
either splits $U$, contains only pendant
vertices, or contains $U$ but not all
vertices of $Z$. In each case its conductance
is strictly greater than $\phi_G(A_1)$.
The centrality of $A_2$ follows similarly
from the completeness of $W$.

On the other hand,
\[
\phi_G(B_1)
=\frac{a+1}{2q(2q-1)+1+a}
\leq a\rho_G(2),
\qquad
\phi_G(B_2)=\rho_G(2).
\]
Thus, $B_1,B_2$ are the subsets that  $a$-approximate  $\rho_G(2)$. 

Finally, since $B_1\triangle A_1=Z$
and $B_2=A_2$, we have that
\[
\frac{\vol(B_1\triangle A_1)}{\vol(A_1)}
=\frac{a}{2q(2q-1)+1+2a}
=\Theta\left(\frac{a}{q^2}\right)
=\Theta\left(\frac{a}{g}\right).
\]
The other permutation gives an even larger
relative symmetric difference, since it
matches the two disjoint cliques.
This completes the proof.
\end{proof}

\section{Cluster Decomposition in Nearly-Linear Time}
\label{sec:fast-GT}
In this section, we present a nearly-linear time algorithm that partitions an input graph into clusters with both the inner and outer conductance guarantees, and prove Theorem~\ref{thm:fast-GT}.

\subsection{Algorithm}

Let $c_0>0$ be the universal constant from Lemma~\ref{lem:fixed11},
and $C_0>0$ be a sufficiently large constant. We define
\begin{equation}\label{eq:def_pstar}
\rho^\star
\triangleq
\min\left\{
\frac{\lambda_k(\mathcal{L}_G)}{10},
C_0k^5\sqrt{\lambda_{k-1}(\mathcal{L}_G)}
\right\},
\qquad
p_\ell
\triangleq
\left(1+\frac1k\right)^\ell\rho^\star .
\end{equation}
Since $\ell\leq k$, we have $p_\ell\leq \mathrm{e}\rho^\star$. We also define
\[
\tau_{\ell} \triangleq \frac{p_{\ell}}{32\sqrt{\mathrm{e}}k^2}.
\]

At a very high level, our designed algorithm consists of $\ell~(1\leq\ell \leq k-1 )$ stages, starting with $\ell=1$ and $A_1=V$. At every stage, the algorithm runs \textsc{Fair-Cut Trimming}~(Algorithm~\ref{alg:fair-trimming}) on every $A_i~(i\in [\ell])$ with $\eps_0\triangleq 1/k$. By Lemma~\ref{lem:backbone-low}, if the input set $A_i~(i\in[\ell])$ satisfies $\phi_G(A_i)\leq p_\ell$, then \textsc{Fair-Cut Trimming} outputs a vertex set $B_i\subseteq A_i$ of positive volume and satisfies 
\[
\phi_G(B_i)
\leq
\left(1+\frac{1}{16k}\right)p_\ell
<
p_{\ell+1}.
\]
Next, the algorithm   runs \textsc{ConstructPartition} on $B_1,\ldots, B_{\ell}$, and obtains a partition $V=P_1\sqcup\cdots\sqcup P_\ell$. At this point, the algorithm runs \textsc{SpectralPartitioning} on every $G\{P_i\}$, and examines the sparse cut $(S_i, P_i\setminus S_i)$ satisfying $\vol(S_i)\leq \vol(P_i)/2$ returned by \textsc{SpectralPartitioning}:
\begin{itemize}
\item If $\phi_{G\{P_i\}}(S_i)\geq \tau_{\ell+1}$ for every $i\in[\ell]$, then every $G\{P_i\}$ has high inner conductance. At this point, 
the algorithm returns $P_1,\ldots, P_{\ell}$ as the output.
\item Otherwise, the algorithm chooses $i\in[\ell]$ satisfying $\phi_{G\{P_i\}}(S_i)< \tau_{\ell+1}$, and examines $\phi_G(S_i\cap B_i)$ and $\phi_G(B_i\setminus (S_i\cap B_i))$:
\begin{itemize}
\item If $\max\{\phi_G(S_i\cap B_i),\phi_G(B_i\setminus (S_i\cap B_i))\}\leq p_{\ell+1}$, then the algorithm uses $S_B\triangleq S_i\cap B_i$ and $T_B \triangleq B_i\setminus (S_i\cap B_i)$ to replace $B_i$ and sets \[(A_1,\ldots,A_{\ell+1})\gets
(B_1,\ldots,B_{i-1},S_B,T_B,B_{i+1},\ldots,B_\ell);\]
\item Otherwise, the algorithm sets \textbf{$(A_1,\ldots,A_{\ell+1})\gets
(B_1,\ldots,B_\ell,S_i\setminus B_i)$.}
\end{itemize}
\end{itemize}
The algorithm increases the value of $\ell$ by $1$, and enters the next stage. When $\ell=k-1$, the algorithm simply runs \textsc{Fair-Cut Trimming} on $A_1,\ldots, A_{\ell}$, and uses the output of
\textsc{ConstructPartition} as the final partition.
See Algorithm~\ref{alg:fast-GT} for a formal description.

We remark that the design of  our algorithm is similar to Algorithm~3 of \cite{GT14} at a very high level: by setting $A_1\leftarrow V$, both algorithms run by stages and perform local refinement on the currently maintained partition in each stage. However, the conditions used in the local refinement in Algorithm~3 of \cite{GT14} are much more involved. As a result, for any input graph $G$ with $n$ vertices and $m$ edges, their algorithm runs in $O(kn\cdot m)$ stages in the worst case. In contrast, our algorithm runs in $O(k)$ stages; this is one of the keys to achieve the nearly-linear running time.

\begin{algorithm}[ht]
	\caption{\textsc{ClusterDecomposition}}
	\label{alg:fast-GT}
	\begin{algorithmic}[1]
		\Require A graph $G$ and an integer $k\geq2$ with $\lambda_k (\mathcal{L}_G) >0$
		\Ensure An $\ell$-partition for some $1\leq\ell<k$
		
		\State Set $\ell\gets1$ and $A_1\gets V$.
	
	\While{true}
	\State Set $p_\ell\gets(1+1/k)^\ell\rho^\star$, $\tau_{\ell}\gets
\frac{p_{\ell}}{32\sqrt \mathrm{e}\,k^2}$
	
	\If{$\ell=k-1$}
	\State Run \textsc{Fair-Cut Trimming}~(Algorithm~\ref{alg:fair-trimming})  on
	$A_1,\ldots,A_\ell$ with $p=p_\ell$ and $\eps=1/2$, obtaining $B_1,\ldots,B_\ell$ 
	\State \Return the output of \textsc{ConstructPartition}~(Algorithm~\ref{alg:fair-attachment}) on
	$B_1,\ldots,B_\ell$.
   \EndIf
	
	\State Run \textsc{Fair-Cut Trimming}~(Algorithm~\ref{alg:fair-trimming}) with $\eps=1/k$
	on every $A_j$, obtaining $B_1,\ldots,B_\ell$.
	
	\State Run \textsc{ConstructPartition}~(Algorithm~\ref{alg:fair-attachment}) on
	$B_1,\ldots,B_\ell$, obtaining
	$V=P_1\sqcup\cdots\sqcup P_\ell$.
	
	\State Run \textsc{SpectralPartitioning} on every $G\{P_i\}$.
	
\State For every $i\in[\ell]$, let $(S_i,P_i\setminus S_i)$ be the cut
returned by \textsc{SpectralPartitioning}, where
$\vol(S_i)\leq\vol(P_i)/2$.

\If{$\phi_{G\{P_i\}}(S_i)\geq\tau_{\ell+1}$ for every $i\in[\ell]$}
\State \Return $P_1,\ldots,P_\ell$.
\EndIf
	
	\State Choose $i\in[\ell]$ such that
$\phi_{G\{P_i\}}(S_i)<\tau_{\ell+1}$, and set $S\gets S_i$.  
	
	\State Replace $S$ by $P_i\setminus S$ if necessary, so that
	$\vol(S\cap B_i)\leq\vol(B_i)/2$.
	
	\State Set $S_B\gets S\cap B_i$,
	$T_B\gets B_i\setminus S_B$, and
	$S_F\gets S\setminus B_i$.
	
	\If{$S_B,T_B$ have positive volume and
		$\max\{\phi_G(S_B),\phi_G(T_B)\}\leq p_{\ell+1}$}
	\State $(A_1,\ldots,A_{\ell+1})\gets
(B_1,\ldots,B_{i-1},S_B,T_B,B_{i+1},\ldots,B_\ell)$.
	\EndIf

      \If{$S_F$ has positive volume and $\phi_G(S_F)\leq p_{\ell+1}$}
	\State $(A_1,\ldots,A_{\ell+1})\gets
(B_1,\ldots,B_\ell,S_F)$.
	\EndIf
	
	\State $\ell\gets\ell+1$.
	\EndWhile
	\end{algorithmic}
\end{algorithm}

\subsection{Analysis}

Now we analyze the performance of Algorithm~\ref{alg:fast-GT}, and prove Theorem~\ref{thm:fast-GT}. 
We first prove the following lemma, which can be viewed as a corollary of Lemma~\ref{lem:one-shot}.

\begin{lemma}\label{lem:local-one-shot} 
	Let $A$ be a vertex set satisfying $\phi_G(A)\leq p$, and   $B$ be the output
    of Algorithm~\ref{alg:fair-trimming} when $A$ is the input and 
	$\eps=\eps_0$. Let $U\subseteq B$, $T=B\setminus U$ and
	$0<\vol(U)<\vol(B)$. If $\max\{\phi_G(U),\phi_G(T)\}\geq(1+\eps_0)p$, then
	\[
	\varphi(U,B)\geq\frac{\eps_0}{4}\qquad and \qquad
     \varphi(T,B)\geq\frac{\eps_0}{4}.
	\]
	Moreover, if $\vol(U)\leq\vol(T)$, then
	\[
	\frac{w(U,T)}{\vol(U)}
	\geq
	\frac{\eps_0}{9}\cdot 
	\max\{\phi_G(U),\phi_G(T)\}.
	\]
\end{lemma}

\begin{proof}
The proof  follows the ones of  Lemma~\ref{lem:one-shot} and Corollary~\ref{cor:relative}; the only difference is that we use the condition $\max\{\phi_G(U),\phi_G(T)\}\geq(1+\eps_0)p$ instead of $\rho_G(k+1)\geq(1+\eps_0)p$ assumed  in Lemma~\ref{lem:one-shot} and Corollary~\ref{cor:relative}.
\end{proof}

Next we show that   whenever \textsc{SpectralPartitioning} returns a cut \(S\subsetneq P_i\) of conductance less than \(\tau_{\ell+1}\), either \(\max\{\phi_G(S_B),\phi_G(T_B)\}\leq p_{\ell+1}\) or \(\phi_G(S_F)\leq p_{\ell+1}\) always holds. Notice that these two events correspond to the two \textsc{if}-conditions in Lines 18 and 21 of Algorithm~\ref{alg:fast-GT}. As a result, the algorithm either splits the current backbone into two vertex sets or adds $S_F$ as a new vertex set.

\begin{lemma}\label{lem:fast-promote}
	Suppose that $A_1,\ldots,A_\ell$ are pairwise disjoint positive-volume sets satisfying $\phi_G(A_j)\leq p_\ell$. We apply Algorithm~\ref{alg:fair-trimming} with $\eps=1/k$ to obtain $B_1,\ldots,B_\ell$, and   apply Algorithm~\ref{alg:fair-attachment} to obtain
	$V=P_1\sqcup\cdots\sqcup P_\ell$. Fix $i\in[\ell]$ and a nontrivial $S\subsetneq P_i$ satisfying $\vol(S\cap B_i)\leq\vol(B_i)/2$, which always holds by replacing $S$ by $P_i\setminus S$ if needed. 
     Set
	$S_B\triangleq S\cap B_i$, $T_B\triangleq B_i\setminus S_B$, and
	$S_F \triangleq S\setminus B_i$. If
	\[
	\frac{w(S,P_i\setminus S)}
	{\min\{\vol(S),\vol(P_i\setminus S)\}}
	<\tau_{\ell+1}=
    \frac{p_{\ell+1}}{32\sqrt{\mathrm{e}}\,k^2},
	\]
	then at least one of the following holds:
	\begin{enumerate}
		\item [\textup{(1)}] $S_B$ and $T_B$ have positive volume and
		$\max\{\phi_G(S_B),\phi_G(T_B)\}\leq p_{\ell+1}$;
		\item [\textup{(2)}] $S_F$ has positive volume and $\phi_G(S_F)\leq p_{\ell+1}$.
	\end{enumerate}
\end{lemma}

\begin{proof}
Suppose that neither statement holds, and we will reach a contradiction by    showing  that
\[
\frac{w(S,P_i\setminus S)}{\vol(S)}\geq \frac{p_{\ell+1}}{32\sqrt{\mathrm{e}}\,k^2}.
\]
	 If $S_B=\emptyset$, then $S=S_F$ and   $\vol(S_F)=\vol(S)>0$. Thus, as (2) doesn't hold, we have $\phi_G(S)>p_{\ell+1}$. Applying Lemma~\ref{lem:attachment} to $Y=S$  gives us that
	\[
   \frac{w(S,P_i\setminus S)}{\vol(S)}\geq \frac{\phi_G(S)}{2\sqrt \mathrm{e}\,k}
	>
	\frac{p_{\ell+1}}{2\sqrt \mathrm{e}\,k},
	\]
    which leads to 
	a contradiction.

    Hence, 
	we can assume that $S_B$ has positive volume. Since $\vol(S_B)\leq\vol(B_i)/2$, the set $T_B$ also has positive volume. As conclusion~(1) fails, we have
	$\max\{\phi_G(S_B),\phi_G(T_B)\}>p_{\ell+1}$. Hence, it holds by Lemma~\ref{lem:local-one-shot} with $\eps_0=1/k$ that 
	\[
	\varphi(S_B,B_i)\geq\frac{1}{4k},\qquad \varphi(T_B,B_i)\geq\frac{1}{4k},\qquad and \qquad \frac{w(S_B,T_B)}{\vol(S_B)}>\frac{p_{\ell+1}}{9k}.
	\]
	If $\vol(S_B)\geq\vol(S_F)$, then $\vol(S)\leq2\vol(S_B)$  and  
	\[
	\frac{w(S,P_i\setminus S)}{\vol(S)}\geq  \frac{w(S_B,T_B)}{2\vol(S_B)}
	\geq
	\frac{p_{\ell+1}}{18k},
	\]
    which again leads to  a contradiction.  
	It remains to consider $\vol(S_F)\geq\vol(S_B)>0$. Since
	$\vol(T_B)\geq\vol(B_i)/2$, $\varphi(S_B,B_i)\geq 1/(4k)$ gives us that 
	\[
	w(S_B,T_B)
	\geq
	\frac{1}{8k}w(S_B,V\setminus B_i)
	\geq
	\frac{1}{8k}w(S_B,S_F).
	\]
	Since $w(S_F,P_i\setminus S_F)\geq w(S_B,S_F)$, we have 
\begin{eqnarray*}
w(S,P_i\setminus S)
&\geq&
w(S_B,B_i\setminus S_B)
+w(S_F,P_i\setminus S_F)
-w(S_B,S_F)\\
&\geq&
w(S_F,P_i\setminus S_F)
-\left(1-\frac{1}{8k}\right)w(S_B,S_F)\\
&\geq&
\frac{1}{8k}w(S_F,P_i\setminus S_F).
\end{eqnarray*}
Since the second statement doesn't hold, we have $\phi_G(S_F)>p_{\ell+1}$. Applying Lemma~\ref{lem:attachment} to $S_F$ gives $w(S_F,P_i\setminus S_F)\geq p_{\ell+1}\vol(S_F)/(2\sqrt \mathrm{e}\,k)$. As $\vol(S)\leq2\vol(S_F)$, we obtain
	\[
	\frac{w(S,P_i\setminus S)}{\vol(S)}
	\geq \frac{p_{\ell+1}}{16\sqrt{\mathrm{e}}\,k^2}\cdot\frac{\vol(S_F)}{\vol(S)} \geq 
	\frac{p_{\ell+1}}{32\sqrt \mathrm{e}\,k^2},
	\]
which leads to a contradiction. Combining the cases above proves the statement. 
\end{proof}

  The following observation will be used in our proof.

\begin{observation}\label{obs:boundedcond}
At the beginning of stage $\ell$, the sets $A_1,\ldots,A_\ell$ are
pairwise disjoint, and have positive volume. Moreover, it holds for every $i\in[\ell]$ that 
$\phi_G(A_i)\leq p_\ell$.
\end{observation}

\begin{proof}
The claim holds initially, since $A_1=V$ and $\phi_G(V)=0$. Suppose it
holds at the beginning of stage $\ell$. If the algorithm performs a refinement, Lines 18 and 21 of the algorithm guarantee that the newly generated subsets also have conductance at most $p_{\ell+1}$. Hence, after increasing $\ell$ by one, all the new subsets satisfy the invariant with parameter $p_{\ell+1}$.
\end{proof}

Now we are ready to prove Theorem~\ref{thm:fast-GT}.

\begin{proof}[Proof of Theorem~\ref{thm:fast-GT}]
We first analyze the running time. We call an entire execution of the \textsc{while-loop}~(Line~2--25) of Algorithm~\ref{alg:fast-GT} a \emph{stage}. Since every non-terminal stage increases
$\ell$ by one and the algorithm terminates once $\ell=k-1$, there are at
most $k-1$ stages. In each non-terminal stage, the algorithm invokes Algorithm~\ref{alg:fair-trimming} with $\eps=1/k$ and Algorithm~\ref{alg:fair-attachment}. By Theorem~\ref{thm:main}, Algorithm~\ref{alg:fair-trimming} runs in $\TildeO(mk)$ time and Algorithm~\ref{alg:fair-attachment} runs in $\TildeO(m\log^2 k)$ time. Moreover, the total running time of \textsc{SpectralPartitioning} at every stage is $\TildeO(m)$, since all the $G\{P_i\}$ are disjoint. Hence, the total running time of Algorithm~\ref{alg:fast-GT} is $\TildeO(mk^2)$.

Next we prove that the subsets returned by the algorithm have the promised guarantee.
Suppose first that the algorithm terminates at some $\ell\leq k-2$. By Line 12 and Lemma \ref{lem:cheeger-cut}, for every $i\in [\ell]$, \textsc{SpectralPartitioning} returns a subset $S_i\subsetneq P_i$ with
\[
\tau_{\ell+1}\leq \phi_{G\{P_i\}}(S_i)\leq\sqrt{2\lambda_2(\mathcal{L}_{G\{P_i\}})}. 
\]
Hence, by Lemma~\ref{lem:cheeger-cut}, we have
\[
\phi(G\{P_i\})\geq \frac{\lambda_2(\mathcal{L}_{G\{P_i\}})}{2}\geq \frac{\tau_{\ell+1}^2}{4}
\]
for every $i\in[\ell]$.

We prove that $\rho^\star=\lambda_k(\mathcal{L}_G)/10$ by contradiction. Suppose that $\rho^\star\neq\lambda_k(\mathcal{L}_G)/10$. Then, by the definition of $\rho^{\star}$~(Equation~\ref{eq:def_pstar}) we have
$\rho^\star=C_0k^5\sqrt{\lambda_{k-1}(\mathcal{L}_G)}$. Since $\ell\leq k-2$, applying Lemma \ref{lem:fixed11} with parameter $k-1$ shows that there exists $i\in[\ell]$ such that
\[
\lambda_2(\mathcal{L}_{G\{P_i\}})
\leq
2c_0(k-1)^6\lambda_{k-1}(\mathcal{L}_G)
\leq
2c_0k^6\lambda_{k-1}(\mathcal{L}_G).
\]
Hence, \textsc{SpectralPartitioning} on $G\{P_i\}$ returns a cut of conductance at most $2\sqrt{c_0}\,k^3\sqrt{\lambda_{k-1}(\mathcal{L}_G)}$. On the other hand, since the algorithm terminates at this stage, the returned cut has conductance at least
\[
\tau_{\ell+1}
=
\frac{p_{\ell+1}}{32\sqrt \mathrm{e}\,k^2}
\geq
\frac{\rho^\star}{32\sqrt \mathrm{e}\,k^2}
=
\frac{C_0}{32\sqrt \mathrm{e}}\,
k^3\sqrt{\lambda_{k-1}(\mathcal{L}_G)}.
\]
Choosing $C_0>64\sqrt{\mathrm{e}c_0}$ gives a contradiction. Hence, 
  $\rho^\star=\lambda_k(\mathcal{L}_G)/10$. This implies that  $\tau_{\ell+1}=\Omega(\lambda_k(\mathcal{L}_G)/k^2)$, and hence
\[
\phi(G\{P_i\})
=
\Omega\left(\frac{\lambda_k^2(\mathcal{L}_G)}{k^4}\right).
\]
Moreover, by the outer-conductance guarantee of Algorithm~\ref{alg:fair-attachment} and Observation~\ref{obs:boundedcond} it holds for every $i\in[\ell]$ that  $\phi_G(P_i)=O(kp_\ell)$. Since $p_\ell\leq \mathrm{e}\rho^\star\leq \mathrm{e}C_0k^5\sqrt{\lambda_{k-1}(\mathcal{L}_G)}$, we have
\[
\phi_G(P_i)
=
O\left(k^6\sqrt{\lambda_{k-1}(\mathcal{L}_G)}\right).
\]

It remains to consider the case in which the algorithm reaches $\ell=k-1$. At this point, $p_{k-1}\leq \mathrm{e}\rho^\star\leq(\mathrm{e}/10)\lambda_k(\mathcal{L}_G)$. Since $\rho_G(k)\geq\lambda_k(\mathcal{L}_G)/2$ and $3\mathrm{e}/20<1/2$, we have
$\rho_G(k)\geq(3/2)p_{k-1}$. Therefore, applying the terminal
partitioning procedure gives
\[
\phi_G(P_i)
=
O\left(k^6\sqrt{\lambda_{k-1}(\mathcal{L}_G)}\right),
\qquad
\phi(G\{P_i\})
=
\Omega\left(\frac{\lambda_k(\mathcal{L}_G)}{k}\right).
\]
This proves the theorem.
\end{proof}

\section{Conclusions}
In this paper we presented a nearly-linear time algorithm  that finds $k$ disjoint subsets under the gap between $\rho_G(k+1)$ and $\rho_G(k)$, improving the prior work~\cite{GT14} which only established  the existence of these subsets. We further presented a nearly-linear time algorithm that finds a $k$-way partition, improving the previously best known algorithm~\cite{GT14}, which runs  in $\Omega(n^3)$ time. Our results constitute the first improvement on this important problem in over a decade. We believe these new results and techniques, like the procedure that boosts an arbitrary $\alpha$-approximation solution for $\rho_G(k)$ to a $(1+\varepsilon)$-approximation solution, will have further applications in graph clustering and related graph problems.



\section*{Acknowledgement}
The authors thank Shayan Oveis Gharan for helpful discussions concerning the issue in the statement of \cite[Lemma 1.1]{GT14}, and in particular for clarifying the role of self-loops in the relevant spectral argument.

\section*{AI Methodology}
We used OpenAI's ChatGPT for feedback on the presentation of the manuscript and for exploring mathematical examples. In particular, the tight example in Proposition \ref{prop:concentrationtight} was suggested by ChatGPT and independently verified by the authors. All proofs in this paper were written and  verified by the authors, who take full responsibility for the correctness and originality of the work.

\makeatletter
\let\original@lbibitem\@lbibitem
\def\@lbibitem[#1]#2{%
  \ifstrequal{#2}{GT14}
    {\original@lbibitem[OGT14]{#2}}
    {\ifstrequal{#2}{LOT14}
       {\original@lbibitem[LOGT14]{#2}}
       {\original@lbibitem[#1]{#2}}}%
}
\makeatother

\bibliographystyle{alphaurl}
\bibliography{refs}

\newpage

\appendix 

\section{Omitted Detail from Section~\ref{sec:pre}\label{sec:pre_app}}\label{apx:sec2}

In this section we present all the proofs omitted from Section~\ref{sec:pre}.

\begin{proof}[Proof of Lemma~\ref{lem:disjoint-rayleigh}]
Since $f_1,\dots,f_k$ have pairwise disjoint supports, they are linearly independent. Hence $U\triangleq \operatorname{span}\{f_1,\dots,f_k\}$ has dimension $k$, and by \eqref{eq:CF} it suffices to show that
$\RQ_G(f)\leq2\max_i\RQ_G(f_i)$ for every nonzero
$f=\sum_{i=1}^k c_i f_i\in U$, where $c_1,\ldots, c_k\in\mathbb{R}$.

By the  disjointness of the supports, we have $f_i(v)f_j(v)=0$ for any $i\neq j$ and therefore
\[
\sum_{v\in V(G)}\deg(v)f(v)^2
=\sum_{v\in V(G)} \deg(v)\left(\sum_{i=1}^kc_if_i(v)\right)^2
=\sum_{i=1}^k c_i^2\sum_{v\in V(G)}\deg(v)f_i(v)^2.
\]
For every non-loop edge $\{u,v\}$, at most two of the quantities
$c_i(f_i(u)-f_i(v))$ are nonzero, and therefore
\[
(f(u)-f(v))^2
\leq
2\sum_{i=1}^k c_i^2(f_i(u)-f_i(v))^2.
\]
Summing over all non-loop edges and dividing by the preceding expression for the denominator gives
\begin{eqnarray*}
\RQ_G(f)&\leq& \frac{\sum_{\substack{\{u,v\}\in E(G) \\ u\neq v}}w(u,v)\sum_{i=1}^k2c_i^2(f_i(u)-f_i(v))^2}{\sum_{i=1}^k c_i^2\sum_{v\in V(G)} \deg(v)f_i(v)^2}\\
&=&\frac{2\cdot\sum_{i=1}^kc_i^2\sum_{\substack{ \{u,v\}\in E(G) \\ u\neq v}}w(u,v)(f_i(u)-f_i(v))^2}{\sum_{i=1}^k c_i^2\sum_{v\in V(G)} \deg(v)f_i(v)^2}\\
&\leq& 2 \max_{1\leq i\leq k} \RQ_G(f_i),
\end{eqnarray*}
which completes the proof.
\end{proof}

\begin{proof}[Proof of Lemma~\ref{lem:LOT}]
It is known by \cite{LOT14} that the same statement holds for any graph in which there is no self-loop on the vertices, and our proof is based on this fact. For the given graph $G=(V,E,w_G)$ with self-loops, we construct a loopless graph $\widehat G$ as follows: starting with $\widehat G=G$, for any $v\in V$ with self-loop we add a new vertex $p_v$ and an edge $\{v,p_v\}$ of weight $w_G(v,v)/2$. Hence, for any vertex $v\in V$ with self-loops we have that 
\[
\deg_{\widehat G}(v)=\deg_G(v)-\frac{w_G(v,v)}{2},
\qquad
\deg_{\widehat G}(p_v)=\frac{w_G(v,v)}{2};
\]
the degrees of the vertices without self-loops are the same in $G$ and $\widehat G$. In particular, $\deg_{\widehat G}(v)+\deg_{\widehat G}(p_v)=\deg_G(v)$ for every vertex $v$
with a self-loop.
 
For any function $f:V(G)\mapsto\mathbb R$, define its extension
$\widehat f:V(\widehat G)\mapsto\mathbb R$ by setting
$\widehat f(v)=f(v)$ for every $v\in V(G)$ and
$\widehat f(p_v)=f(v)$ for every $p_v\in V(\widehat G)\setminus V(G)$. Since $\deg_{\widehat G}(v)\widehat f(v)^2+
\deg_{\widehat G}(p_v)\widehat f(p_v)^2=\deg_G(v)f(v)^2$, we have 
\[
\RQ_{\widehat G}(\widehat f)=\RQ_G(f).
\]

Let $f_1,\ldots,f_k$ be the harmonic eigenfunctions corresponding to
$\lambda_1(\mathcal{L}_G),\ldots,\lambda_k(\mathcal{L}_G)$. Since the extension
$f\mapsto\widehat f$ is linear and injective, the space spanned by 
$ \widehat f_1,\ldots,\widehat f_k$ has
dimension $k$. Hence, it holds by \eqref{eq:CF} that 
\[
\lambda_k(\mathcal{L}_{\widehat G})
\leq
\max_{\widehat f\in
\operatorname{span}\{\widehat f_1,\ldots,\widehat f_k\}\setminus\{0\}}
\RQ_{\widehat G}(\widehat f)
=
\max_{f\in
\operatorname{span}\{f_1,\ldots,f_k\}\setminus\{0\}}
\RQ_G(f)
=
\lambda_k(\mathcal{L}_G).
\]
Applying the loopless version of the statement~\cite{LOT14} to $\widehat G$, we obtain
$k$ nonzero functions
$\widehat g_1,\ldots,\widehat g_k$ with pairwise disjoint supports such
that
\[
\RQ_{\widehat G}(\widehat g_i)
\leq
c k^6\lambda_k(\mathcal{L}_{\widehat G})
\leq
c k^6\lambda_k(\mathcal{L}_G)
\qquad\text{for every }i\in[k];
\]
here $c$ is a universal constant.
 
Let $\alpha\triangleq c k^6\lambda_k(\mathcal{L}_G)$. Our analysis is based on case distinction. 

\emph{Case~1~($\alpha\geq1/4$):} we choose any $k$
distinct vertices $v_1,\ldots,v_k\in V(G)$ and let
$g_i=\mathbf 1_{v_i}$, where $\mathbf 1_v(v)\triangleq 1$ and $\mathbf{1}_v (u)\triangleq 0$ for any $u\neq v$. By construction, $g_1,\ldots, g_k$ have pairwise disjoint supports, and  $\RQ_G(g_i)\leq1\leq4\alpha$. Hence, the statement follows in the case.

\emph{Case~2~($\alpha<1/4$):} For any $i\in[k]$, let $g_i$ be the restriction of $\widehat g_i$ to $V(G)$, and by construction $g_1,\ldots, g_k$ have pairwise disjoint supports.  We will prove  that
\[
\RQ_G(g_i)\leq3\RQ_{\widehat G}(\widehat g_i),
\]
which will follow if we show that 
$g_i^{\top}D_Gg_i\geq
\widehat g_i^{\top}D_{\widehat G}\widehat g_i/3$.
Indeed, for any vertex $v$ with a self-loop, since
$\widehat g_i(p_v)^2\leq
2\widehat g_i(v)^2+
2(\widehat g_i(p_v)-\widehat g_i(v))^2$
and
$w_{\widehat G}(v,p_v)=w_G(v,v)/2\leq \deg_G(v)/2$, we have
\begin{align*}
\widehat g_i^{\top}D_{\widehat G}\widehat g_i
&=
\sum_{v\in V(G)}
\left(\deg_G(v)-\frac{w_G(v,v)}{2}\right)\widehat g_i(v)^2
+\sum_{v\in V(G)}
\frac{w_G(v,v)}{2}\widehat g_i(p_v)^2\\
&\leq
\sum_{v\in V(G)}
\left(\deg_G(v)+\frac{w_G(v,v)}{2}\right)
\widehat g_i(v)^2
+2\sum_{v\in V(G)}
w_{\widehat G}(v,p_v)
(\widehat g_i(p_v)-\widehat g_i(v))^2\\
&\leq
\frac32\cdot g_i^{\top}D_Gg_i
+2\alpha\cdot
\widehat g_i^{\top}D_{\widehat G}\widehat g_i\\
&\leq
\frac32\cdot g_i^{\top}D_Gg_i
+\frac12\cdot
\widehat g_i^{\top}D_{\widehat G}\widehat g_i,
\end{align*}
which implies that 
\[
\widehat g_i^{\top}D_{\widehat G}\widehat g_i
\leq
3\cdot g_i^{\top}D_Gg_i.
\]
Hence, it holds that 
\[
\RQ_G(g_i)
\leq
3\RQ_{\widehat G}(\widehat g_i)
\leq
3\alpha.
\]
Combining the two cases,   there exist $k$ disjointly supported nonzero functions on $G$ whose Rayleigh quotients are at most
$4c k^6\lambda_k(\mathcal{L}_G)$. Absorbing the factor $4$ into the universal constant $c_0$ proves the statement.
\end{proof}

\begin{proof}[Proof of Lemma~\ref{lem:fixed11}]
Let $g_1,\dots,g_k$ be the vectors obtained from Lemma~\ref{lem:LOT}, and $g_{i,j}$   the restriction of $g_i$ to $P_j$. Hence,
\begin{eqnarray*}
\RQ_G(g_i)
&=&
\frac{\sum_{\substack{ \{u,v\}\in E(G) \\ u\neq v}}w(u,v)(g_i(u)-g_i(v))^2}
{\sum_{v\in V(G)}\deg(v)g_i(v)^2}\\
&\geq&
\frac{\sum_{j=1}^\ell
\sum_{\substack{\{u,v\}\in E(G)\\u\neq v, u,v\in P_j}}
w(u,v)(g_i(u)-g_i(v))^2}
{\sum_{j=1}^\ell\sum_{v\in P_j}\deg(v)g_i(v)^2}\\
&\geq&
\min_{\substack{1\leq j\leq\ell\\g_{i,j}\neq 0}}
\frac{
\sum_{\substack{\{u,v\}\in E(G)\\ u\neq v, u,v\in P_j}}
w(u,v)(g_i(u)-g_i(v))^2}
{\sum_{v\in P_j}\deg(v)g_i(v)^2}
=
\min_{\substack{1\leq j\leq\ell\\g_{i,j}\neq 0}}
\RQ_{G\{P_j\}}(g_{i,j}).
\end{eqnarray*}
For each $i\in[k]$,  we choose $j(i)\in[\ell]$ such that
$g_{i,j(i)}\neq0$ and
\[
\RQ_{G\{P_{j(i)}\}}(g_{i,j(i)})
\leq
\RQ_G(g_i).
\]
As $P_1,\dots,P_\ell$ form a partition of $V$ and
$\ell\leq k-1$, by the pigeonhole principle there exist
$j^*\in[\ell]$ and distinct $s,t\in[k]$ such that
$j(s)=j(t)=j^*$. In particular, $g_{s,j^*}$ and $g_{t,j^*}$ are nonzero and have disjoint supports. Moreover, it holds that 
\[
\RQ_{G\{P_{j^*}\}}(g_{s,j^*})\leq\RQ_G(g_s),
\qquad
\RQ_{G\{P_{j^*}\}}(g_{t,j^*})\leq\RQ_G(g_t).
\]
Thus, by Lemma~\ref{lem:disjoint-rayleigh} and Lemma~\ref{lem:LOT}, we have
\[
\lambda_2(\mathcal{L}_{G\{P_{j^*}\}})
\leq
2\max\left\{
\RQ_{G\{P_{j^*}\}}(g_{s,j^*}),
\RQ_{G\{P_{j^*}\}}(g_{t,j^*})
\right\}
\leq
2\max\{\RQ_G(g_s),\RQ_G(g_t)\}
\leq
2c_0k^6\lambda_k(\mathcal{L}_G),
\]
which completes the proof.
\end{proof}

\section{A Counterexample for Lemma~1.1 of \cite{GT14}}\label{sec:cex}\label{apx:counterexamples}

In this subsection, we present a counterexample to show that the original statement of Lemma~1.1 in \cite{GT14} is incorrect. However, we also note that their results can be saved by proving a similar bound for $G\{P_i\}$s (see Lemma \ref{lem:fixed11}) and replacing induced subgraphs by degree-preserving induced subgraphs everywhere in their proofs. 

\begin{lemma}[{Lemma 1.1 of \cite{GT14}}]\label{lem:GT11}
There is a universal constant $c_0>0$ such that for any $k\ge 2$ and any partitioning of $V$ into $\ell\le k-1$ sets $P_1,\dots,P_\ell$,
\[
\min_{1\le i\le \ell}\lambda_2(\mathcal{L}_{G[P_i]}) \;\le\; 2c_0k^6\lambda_k ,
\]
where $\lambda_2(\mathcal{L}_{G[P_i]})$ is the second eigenvalue of the normalized Laplacian matrix of the induced graph $G[P_i]$.
\end{lemma}
 
Fix an integer $d\ge 3$ and a constant $h\in(0,1]$ such that for infinitely many $n$ there is a $d$-regular graph $X$ on $n$ vertices with
\begin{equation}\label{eq:expander}
e_X\bigl(A,V(X)\setminus A\bigr)\;\ge\;h\,d\cdot\min\{|A|,\,n-|A|\}
\qquad\text{for all }A\subseteq V(X);
\end{equation}
such families exist; one may consider random $d$-regular graphs or explicit expanders \cite{HLW06}.
 
\begin{figure}[ht]
\centering
\begin{tikzpicture}[
  vtx/.style={circle, fill=black, inner sep=1.4pt},
  every node/.style={font=\small}
]
\draw[thick, fill=black!5] (0,0) ellipse (1.4 and 2.4);
\node at (-0.25,0) {$X_1$};
\draw[thick, fill=black!5] (8.4,0) ellipse (1.4 and 2.4);
\node at (8.65,0) {$X_2$};
 
\node[vtx, label={[label distance=-2pt]below left:{$x_1$}}] (x1) at (1.05,1.6) {};
\node[vtx, label={[label distance=-2pt]below left:{$x_2$}}] (x2) at (1.05,0.7) {};
\node[vtx, label={[label distance=-2pt]above left:{$x_m$}}] (xm) at (1.05,-1.6) {};
\node at (1.05,-0.45) {$\vdots$};
 
\node[vtx, label=above:{$t_1$}] (t1) at (3.1,1.6) {};
\node[vtx, label={[label distance=-1pt]left:{$t_2$}}] (t2) at (3.1,0.7) {};
\node[vtx, label=below:{$t_m$}] (tm) at (3.1,-1.6) {};
\node at (3.1,-0.45) {$\vdots$};
 
\node[vtx, label=above:{$t'_1$}] (s1) at (5.3,1.6) {};
\node[vtx, label={[label distance=-1pt]right:{$t'_2$}}] (s2) at (5.3,0.7) {};
\node[vtx, label=below:{$t'_m$}] (sm) at (5.3,-1.6) {};
\node at (5.3,-0.45) {$\vdots$};
 
\node[vtx, label={[label distance=-2pt]below right:{$y_1$}}] (y1) at (7.35,1.6) {};
\node[vtx, label={[label distance=-2pt]below right:{$y_2$}}] (y2) at (7.35,0.7) {};
\node[vtx, label={[label distance=-2pt]above right:{$y_m$}}] (ym) at (7.35,-1.6) {};
\node at (7.35,-0.45) {$\vdots$};
 
\foreach \a in {t1,t2,tm}{\foreach \b in {s1,s2,sm}{\draw[gray!70] (\a)--(\b);}}
 
\draw[thick] (x1)--(t1); \draw[thick] (x2)--(t2); \draw[thick] (xm)--(tm);
\draw[thick] (y1)--(s1); \draw[thick] (y2)--(s2); \draw[thick] (ym)--(sm);
 
\draw[dashed, thick] (4.2,-2.9) -- (4.2,2.2);
\node at (4.2,2.6) {$K_{m,m}$};
\node at (1.3,-3.15) {$P_1=V(X_1)\cup T$};
\node at (7.1,-3.15) {$P_2=V(X_2)\cup T'$};
\end{tikzpicture}
\caption{The graph $G_n$ and the partition $V(G_n)=P_1\cup P_2$ (dashed line); only three of the $m$ vertices of each of $T$ and $T'$ are drawn. The sets $T$ and $T'$ induce a complete bipartite graph, and $t_i$ and $t'_i$ are joined to the hosts $x_i\in V(X_1)$ and $y_i\in V(X_2)$, which are distinct as $i$ varies. Every $T$--$T'$ edge is cut by the partition, so $G_n[P_1]$ is $X_1$ with $m$ pendant vertices attached at distinct hosts, and likewise $G_n[P_2]$ is $X_2$ with $m$ pendant vertices attached at distinct hosts.}
\label{fig:Gn}
\end{figure}
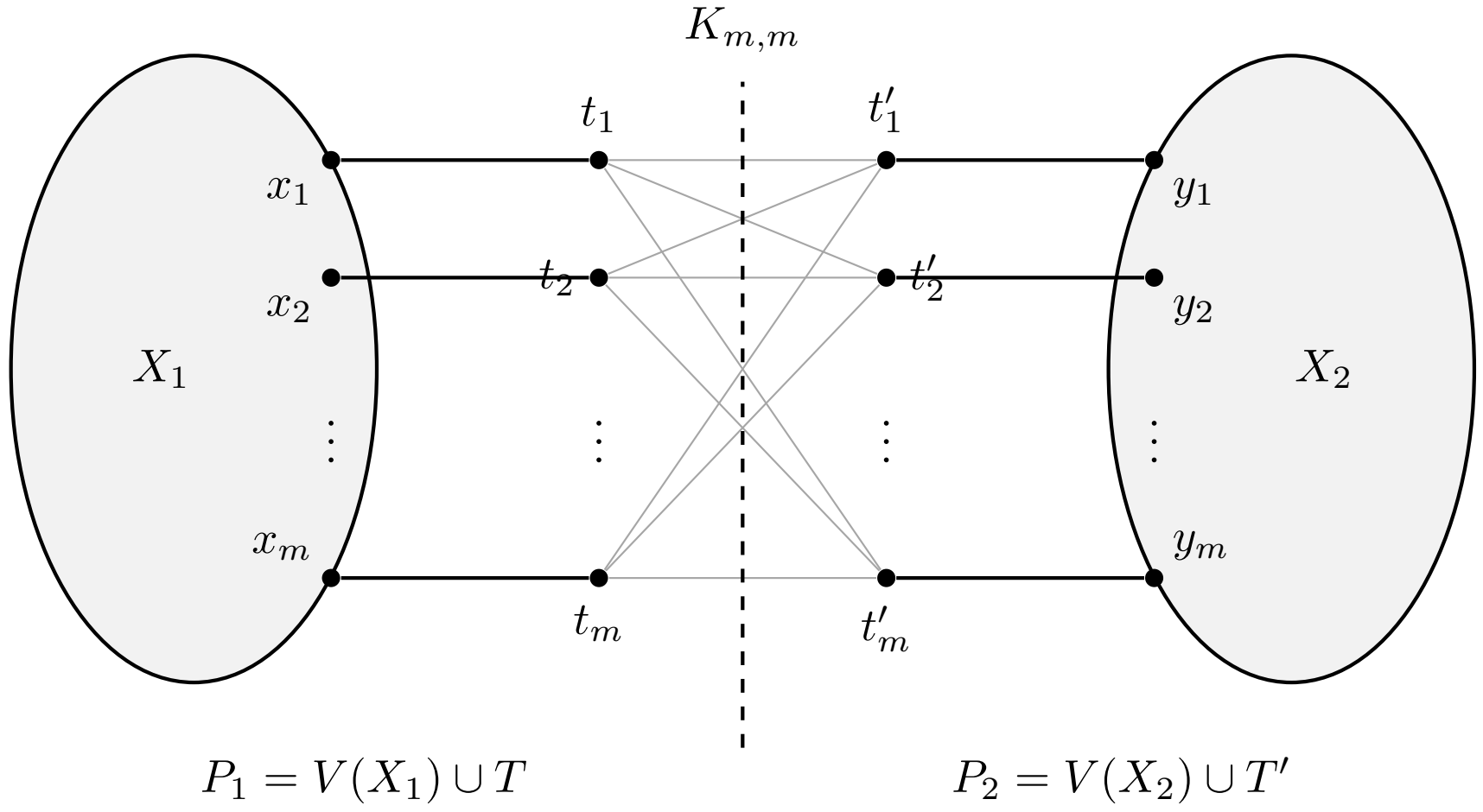

\begin{definition}[the graph $G_n$]\label{def:Gn}
Let $X_1,X_2$ be disjoint copies of $X$ satisfying \eqref{eq:expander} on $n$ vertices, let $m:=\lfloor\sqrt n\rfloor\le n$, and let $T=\{t_1,\dots,t_m\}$ and $T'=\{t'_1,\dots,t'_m\}$ be sets of new vertices. The (unweighted) graph $G_n$ on $V=V(X_1)\cup V(X_2)\cup T\cup T'$ has edge set
\[
E(X_1)\;\cup\;E(X_2)\;\cup\;\{t_it'_j:\,1\le i,j\le m\}\;\cup\;\{t_ix_i:\,1\le i\le m\}\;\cup\;\{t'_iy_i:\,1\le i\le m\},
\]
where $x_1,\dots,x_m$ are distinct vertices of $X_1$ and $y_1,\dots,y_m$ distinct vertices of $X_2$. Set $P_1:=V(X_1)\cup T$ and $P_2:=V(X_2)\cup T'$.
\end{definition}
Thus, each $t\in T\cup T'$ has degree $m+1$ in $G_n$, each vertex in $\{x_1,x_2,\dots,x_m\}\cup \{y_1,y_2,\dots, y_m\}$ has degree $d+1$, and all other vertices have degree $d$; see Figure~\ref{fig:Gn}.
 
We will need to use the following lemma.

\begin{lemma}[{\cite[Proposition~1.1]{GT14}}]\label{fact:disjoint}
Let $H$ be a weighted graph and let $f_1,\dots,f_k$ be nonzero functions on $V(H)$ with pairwise disjoint supports. Then
\[
\lambda_k(\mathcal{L}_H)\;\le\; 2\max_{1\le i\le k}\RQ_H(f_i).
\]
\end{lemma}
 
\begin{proposition}\label{prop:lam3}
$\lambda_3(\mathcal{L}_{G_n})\le 2n^{-1/2}$.
\end{proposition}
 
\begin{proof}
The sets $A_1=V(X_1)$, $A_2=V(X_2)$, $A_3=T\cup T'$ are disjoint, and for an indicator, $\RQ(\one_A)=w(A\to V)/\vol(A)=\phi(A)$. As the only edges leaving $A_1$ are the $m$ $t_ix_i$s and $\vol(A_1)=dn+m$, $\RQ(\one_{A_1})=m/(dn+m)$ and similarly $\RQ(\one_{A_2})=m/(dn+m)$. Since the only edges leaving $A_3$ are the $2m$ matching edges ($t_ix_is$ and $t_i'y_i$s) and $\vol(A_3)=2m(m+1)$, $\RQ(\one_{A_3})=1/(m+1)$. By Lemma~\ref{fact:disjoint} with $k=3$,
\[
\lambda_3(\mathcal{L}_{G_n})\le 2\max\Bigl\{\frac{m}{dn+m},\,\frac{1}{m+1}\Bigr\}=\frac{2}{m+1}\le  \frac{2}{\sqrt n},
\]
where we use $m(m+1)\le dn+m$ (which holds since $m^2\leq n$) for the first term and $m+1\ge\sqrt n$ for the second. This completes the proof.
\end{proof}
 
\begin{proposition}\label{prop:induced}
Let $c_*:=\min\bigl\{\tfrac{1}{d+3},\,\tfrac{hd}{2(d+3)}\bigr\}>0$. For all sufficiently large $n$,
\[
\phi\bigl(G_n[P_1]\bigr)\ \ge\ c_*
\qquad\text{and hence}\qquad
\lambda_2\bigl(\mathcal{L}_{G_n[P_1]}\bigr)\ \ge\ c_*^2/2,
\]
and the same holds for $P_2$.
\end{proposition}
 
\begin{proof}
We only prove the result for $P_1$ as the argument for $P_2$ is identical. Recall that $G_n[P_1]$ is $X_1$ together with $m$ pendant vertices $t_1, \dots,t_m$ attached to $x_1,\dots,x_m$. In particular, $\vol_{G_n[P_1]}(P_1)=dn+2m$.
 
Let $\emptyset\ne S\subsetneq P_1$ with $\vol_{G_n[P_1]}(S)\le \vol_{G_n[P_1]}(P_1)/2=(dn+2m)/2$, and let $A:=S\cap V(X_1)$ and $B:=S\cap T$. Then $\vol_{G_n[P_1]}(S)\le(d+1)|A|+|B|$, and one can observe that the cut $w_{G_n[P_1]}(S\to P_1)$ is at least $e_{X_1}(A,V(X_1)\setminus A)$ and also at least $|B|-|A|$.
 
If $|B|\ge 2|A|$, then the cut $w_{G_n[P_1]}(S\to P_1)$ has at least $|B|-|A|\ge |B|/2$ edges, and $\vol_{G_n[P_1]}(S)\le(d+1)|A|+|B|\leq (d+1)\frac{|B|}{2}+|B|=\tfrac{d+3}{2}|B|$, so $\phi_{G[P_1]}(S)\ge\tfrac{1}{d+3}$.
 
If $|B|<2|A|$, then $d|A|\le\vol_{G_n[P_1]}(S)\le(dn+2m)/2$ and therefore $|A|\le n/2+m/d\le 2n/3$ for $n$ large (as $m=O(\sqrt n)$). Thus, $n-|A|\geq \frac{|A|}{2}$ and hence $\min\{|A|,n-|A|\}\ge |A|/2$. By \eqref{eq:expander} the cut is at least $hd|A|/2$, while $\vol_{G_n[P_1]}(S)\le(d+1)|A|+|B|\le(d+1)|A|+2|A|=(d+3)|A|$, so $\phi_{G_n[P_1]}(S)\ge\tfrac{hd}{2(d+3)}$.
 
Thus, $\phi(G_n[P_1])\ge c_*$ and therefore by Cheeger's inequality, $\lambda_2(\mathcal{L}_{G_n[P_1]})\ge\phi(G_n[P_1])^2/2\ge c_*^2/2$.
\end{proof}
 
Combining Propositions \ref{prop:lam3} and \ref{prop:induced} gives the following result, which in particular implies Lemma~\ref{lem:GT11} is false.
 
\begin{theorem}\label{thm:cex}
 For all sufficiently large $n$, $G_n$ and the partition $V(G_n)=P_1\cup P_2$ defined in Definition~\ref{def:Gn} satisfy
\[
\min_{1\leq i\leq 2}\lambda_2\bigl(\mathcal{L}_{G_n[P_i]}\bigr)\ \ge\ c_*^2/2\ =\ \Omega(1),
\qquad\text{while}\qquad
2c_0\cdot 3^6\cdot\lambda_3(\mathcal{L}_{G_n})= O(n^{-1/2})\ \xrightarrow[n\to\infty]{}\ 0 .
\]
Thus, Lemma~\ref{lem:GT11} is false. Moreover, there is no exponent $\gamma>0$ such that
$\min_{1\le i\le2}\lambda_2(\mathcal{L}_{G[P_i]})\le O(\,\lambda_3(\mathcal{L}_G)^{\gamma})$
holds for all graphs $G$.
\end{theorem}
 
\begin{remark}
    In fact, there are similar counterexamples of Lemma \ref{lem:GT11} for every $k\geq3$.
\end{remark}
\end{document}